\RequirePackage{fix-cm}
\documentclass[onecollarge]{svjour3}
\smartqed  
\usepackage{xspace}
\usepackage{graphicx}
\usepackage{amssymb,amsbsy,amsmath}
\usepackage{enumitem}
\usepackage{lineno}
\usepackage{hyperref}
\usepackage{color}
\modulolinenumbers[5]
\newcommand{\nc}[1]{\newcommand #1}
\newcommand{\rnc}[1]{\renewcommand #1}
\nc{\etal}{et~al.\xspace}
\nc{\red}[1]{\textcolor{red}{#1}}
\nc{\myquote}[1]{`#1'}
\nc{\x}[1]{\mbox{#1}}
\rnc{\matrix}[2]{\left[\!\!\begin{array}{#1} #2\end{array}\!\!\right]}
\rnc{\vector}[1]{\matrix{c}{#1}}
\nc{\suml}[2]{\sum \limits_{#1}^{#2}}
\nc{\real}[1]{\Re\lbrace #1 \rbrace}
\nc{\imag}[1]{\Im\lbrace #1 \rbrace}
\nc{\conj}{\overline}
\nc{\g}[1]{\x{$#1$}}
\nc{\e}[2]{\begin{equation} #1 \label {eq:#2} \end{equation}}
\nc{\eal}[2]{\begin{equation} \begin{aligned} #1 \label {eq:#2} \end{aligned} \end{equation}}
\nc{\ea}[2]{\begin{eqnarray}
#1 \label {eq:#2}
\end{eqnarray}}
\nc{\inv}{^{-1}}
\nc{\tra}{^{\mathrm T}}
\nc{\herm}{^{\mathrm H}}
\nc{\fabstand}{\,}
\nc{\fp}{\fabstand.}
\nc{\fk}{\fabstand,}
\nc{\mm}[1]{\mathbf{#1}}
\nc{\mms}[1]{\boldsymbol{#1}}
\nc{\ie}{i.\,e.\xspace}
\nc{\eg}{e.\,g.\xspace}
\nc{\cf}{cf.\xspace}
\nc{\dd}{{\mathrm{d}}}
\nc{\ii}{{\mathrm{i}}}
\nc{\jj}{\ii}
\rnc{\phi}{\varphi}
\nc{\ee}{{\mathrm{e}}}
\nc{\fex}{f_{\mathrm{ex}}}
\nc{\fref}[1]{\x{Fig.~\ref{fig:#1}}}
\nc{\frefs}[1]{\x{Figs.~\ref{fig:#1}}}
\nc{\eref}[1]{\x{Eq.~(\ref{eq:#1})}}
\nc{\erefs}[1]{\x{Eqs.~(\ref{eq:#1})}}
\nc{\erefo}[1]{(\ref{eq:#1})}
\nc{\tref}[1]{\x{Tab.~\ref{tab:#1}}}
\nc{\sref}[1]{\x{section~\ref{sec:#1}}}
\nc{\srefo}[1]{\ref{sec:#1}}
\nc{\srefs}[1]{\x{sections~\ref{sec:#1}}}
\nc{\ssref}[1]{\x{subsection~\ref{sec:#1}}}
\nc{\aref}[1]{\x{Appendix~\ref{asec:#1}}}
\nc{\fig}[3][tbh]{
\begin{figure}[#1]
\centering
\includegraphics{figures/#2}
\caption{#3\label{fig:#2}}
\end{figure}}
\nc{\figc}[3][tbh]{
\begin{figure}[#1]
\centering
\includegraphics[width=0.5\columnwidth]{figures/#2}
\caption{#3\label{fig:#2}}
\end{figure}}
\nc{\figw}[3][tbh]{
\begin{figure*}[#1]
\centering
\includegraphics[width=1.0\textwidth]{figures/#2}
\caption{#3\label{fig:#2}}
\end{figure*}}
\nc{\fighw}[3][tbh]{
\begin{figure*}[#1]
\centering
\includegraphics[width=0.5\textwidth]{figures/#2}
\caption{#3\label{fig:#2}}
\end{figure*}}
\nc{\figs}[4][tbh]{
\begin{figure*}[#1]
\centering
\includegraphics[width=#4\textwidth]{figures/#2}
\caption{#3\label{fig:#2}}
\end{figure*}}
\nc{\tab}[5][tbh]{\begin{table}[#1]\centering\caption{#4\label{tab:#5}}\vspace{0.5cm}\begin{tabular}{#2}\hline #3 \\ \hline\end{tabular}\end{table}}
\journalname{Archive of Applied Mechanics}
\begin{document}

\title{Towards understanding the self-adaptive dynamics of a harmonically forced beam with a sliding mass} 

\titlerunning{Self-adaptive dynamics of a beam with a sliding mass}        

\author{Malte Krack \and Noha Aboulfotoh \and Jens Twiefel \and J\"org Wallaschek \and Lawrence A. Bergman \and Alexander F. Vakakis}


\institute{M. Krack \at
              Institute of Aircraft Propulsion Systems, University of Stuttgart, Pfaffenwaldring 6, 70569 Stuttgart, Germany, \email{malte.krack@ila.uni-stuttgart.de}
            \and
            N. Aboulfotoh, J. Twiefel, J. Wallaschek \at
            Institute of Dynamics and Vibration Research, Leibniz Universit\"at Hannover, Appelstr. 11, Hannover, Germany
            \and
            L. A. Bergman \at
              Department of Aerospace Engineering, University of Illinois at Urbana-Champaign, 104 S. Wright Street, Urbana, IL 61801, USA
            \and
            A. F. Vakakis \at
              Department of Mechanical Science and Engineering, University of Illinois at Urbana-Champaign, 1206 W. Green Street, Urbana, IL 61801, USA
}

\date{Received: date / Accepted: date}

\maketitle

\begin{abstract}
A mechanical system consisting of an elastic beam under harmonic excitation and an attached sliding body is investigated. Recent experimental observations suggest that the system passively (self-)adapts the axial location of the slider to achieve and maintain a condition of self-resonance, which could be useful in applications such as energy harvesting. The purpose of this work is to provide a theoretical explanation of this phenomenon based on an appropriate model. A key feature of the proposed model is a small clearance between the slider and the beam. This clearance gives rise to backlash and frictional contact interactions, both of which are found to be essential for the self-adaptive behavior. Contact is modeled in terms of the Coulomb and Signorini laws, together with the Newton impact law. The set-valued character of the contact laws is accounted for in a measure differential inclusion formulation. Numerical integration is carried out using Moreau's time stepping scheme. The proposed model reproduces qualitatively most experimental observations. However, although the system showed a distinct self-adaptive character, the behavior was found to be non-resonant for the considered set of parameters. Beside estimating the relationship between resonance frequency and slider location, the model permits predicting the operating limits with regard to excitation level and frequency. Finally, some specific dynamical phenomena such as hysteresis effects and transient resonance captures underline the rich dynamical behavior of the system.
\end{abstract}

\section{Introduction\label{sec:intro}}
Self-adaptive systems have the special ability to passively adjust their dynamical characteristics, depending on certain operating conditions. Two of the most important examples are self-resonant and self-damping systems. An self-resonant system tunes itself onto the frequency of an applied harmonic excitation, and thus achieves and maintains resonance over a wide range of excitation frequencies. In contrast, a self-damping system detunes its natural frequencies away from the excitation frequency, and thus avoids resonance coincidence, or even achieves anti-resonance in a wide frequency band. From a theoretical point of view, it is important to note that self-adaptive systems are \textit{a priori} nonlinear. The utilization of self-adaptive behavior has several advantages over active means of (de)tuning, including the independence of additional electronic components and a power supply \cite{Aboulfotoh.2013}. Self-adaptive systems have a wide range of potential applications in engineering. The self-damping behavior can be useful for vibration suppression, for instance, to ensure structural integrity or to reduce noise. The self-resonant behavior can be exploited, \eg, for broadband energy harvesting \cite{Tang.2010,Twiefel.2013}. Since the self-adaptation is achieved by (de)tuning, self-resonant and self-damping are well-suited in an environment where the relevant excitation spectrum exhibits a dominant, time-variant frequency. This is the load scenario on which we will focus throughout this work. Such a situation is typical for, \eg, variable-speed rotating machinery. As opposed to this, self-adaptive systems are less suitable in environments with random broadband excitation.
\\
Self-adaptive behavior has been reported both for fluids \cite{Brazovskaia.1998,Boudaoud.1999b} and for solid systems. We will focus on solid systems in this work. A solid system can be (de)tuned by changing its geometry, boundary conditions or its prestress. Various means have been developed for the active adjustment of these properties, see \eg \cite{Zhu.2010,Huang.2012}. The passive adjustment capabilities of these properties is far less developed and understood. So far, self-adaptation has mainly been reported for either a string \cite{Boudaoud.1999}, a beam \cite{Babitsky.1993,Thomsen.1996,Miranda.1998,Miller.2013} or a rigid plate \cite{Wang.2014} with an attached, self-arranging mass. In these systems, the attached mass can assume an equilibrium location along the base system's axis at steady state. This steady location influences the natural frequencies of the system. The equilibrium position may depend on the frequency of the applied harmonic excitation. Thus, the system can passively (de)tune with respect to the excitation.
\\
The theoretical explanation of the self-adaptive behavior is based on similar models in the aforementioned references \cite{Boudaoud.1999,Babitsky.1993,Thomsen.1996,Miranda.1998,Miller.2013,Wang.2014}. The wave equation was used to describe the lateral vibrations of the string in \cite{Boudaoud.1999}, while the Euler-Bernoulli theory was applied to describe the bending vibrations of the beam in \cite{Babitsky.1993,Thomsen.1996,Miranda.1998,Miller.2013}. The slider is commonly considered as point mass or as a rigid body. In the aforementioned references, the interface between base system and sliding mass was always assumed as perfect, \ie, without clearance. To account for frictional effects between the slider and the beam, the Coulomb law was incorporated into the model in \cite{Miranda.1998}. The driving force of the slider can be explained by the projection of the transverse acceleration of the slider onto the deformed base system. It can be shown that the slider's equilibrium points are usually the nodes and the anti-nodes of the vibration deflection shapes. The stability of the equilibrium points may change, depending on rotating inertia effects \cite{Miranda.1998}.
\\
The work of Miller \etal \cite{Miller.2013,Pillatsch.2013} provided the primary motivation of the present study. They provide one of the very few successful experimental demonstrations of self-adaptive behavior in a mechanical system. Their basic setup consists of a clamped-clamped elastic beam carrying a sliding mass under harmonic base excitation. Miller \etal show that their setup can be scaled and operated in a relatively wide range. More specifically, they considered three different variants with nominal frequencies of $45-140~\mathrm{Hz}$, and observed self-resonant behavior in a frequency bandwidth of $6-40~\mathrm{Hz}$ and excitation levels ranging between $0.007$ and $2~g$ acceleration amplitude. They focused on the frequency range around the first bending mode of each beam. The slider assumed equilibrium positions away from the center and the clamping. It was noted in \cite{Pillatsch.2013} that the clearance between the slider and the beam has a crucial influence on the self-adaptive behavior. If the clearance is too tight, the mass does not move and if it is too large, the mass would start rattling and dissipating energy. There is also a video available of their experiment where the self-resonant behavior is shown \cite{Pillatsch.2013b}.
\\
Assuming that the vibrational deflection shape remains close to that of the first beam mode (without or with bonded slider), the equilibrium positions observed in \cite{Miller.2013,Pillatsch.2013} are neither anti-nodes nor nodes. This behavior cannot be explained with the aforementioned modeling approach, which suggests that a crucial effect is not captured. The purpose of the present study is to develop an appropriate model that captures the essential experimental observations. 
An important goal is to assess the importance of unilateral and frictional contact interactions due to the small clearance between the slider and the beam. Moreover, a definition of resonance is developed, which lays the foundation for estimating the location-dependent resonance frequency. This represents the basis for a theoretical explanation of the self-adaptive behavior encountered by Miller \etal.
\\
The remaining article is organized as follows. In \sref{modeling_dynamic_analysis}, the model and dynamical analysis procedure are presented. In \sref{typical}, the typical dynamical behavior is presented in detail, aiming at providing insight into the local and global dynamics during the adaptation phase as well as at steady state. In \sref{parameter_influences}, the influences of clearance, frictional effects and excitation level are shown. In \sref{steady}, the steady-state operating behavior, including its limits, is addressed. In \sref{resonance}, the notion of resonance is defined for the invariant system and a method for approximating the location-dependent resonance frequency is developed. In \sref{unsteady}, the effects of unsteady operation are explored. This paper ends with the conclusions in \sref{conclusions}.

\section{Modeling and simulation\label{sec:modeling_dynamic_analysis}}
In this section, a model of the self-adaptive system is developed. In general, it is assumed that for in-plane oscillations, the behavior of the system can be captured properly with a two-dimensional model. In \ssref{decoupled}, a description of the underlying components in the absence of contact interactions is presented. The contact modeling is addressed in \ssref{contact}. The computational method that was applied for the dynamical analysis in this work is mentioned in \ssref{time_integration}. Finally, the nominal model parameters are presented and discussed in \ssref{nominal_parameters}.
\figw[tbh]{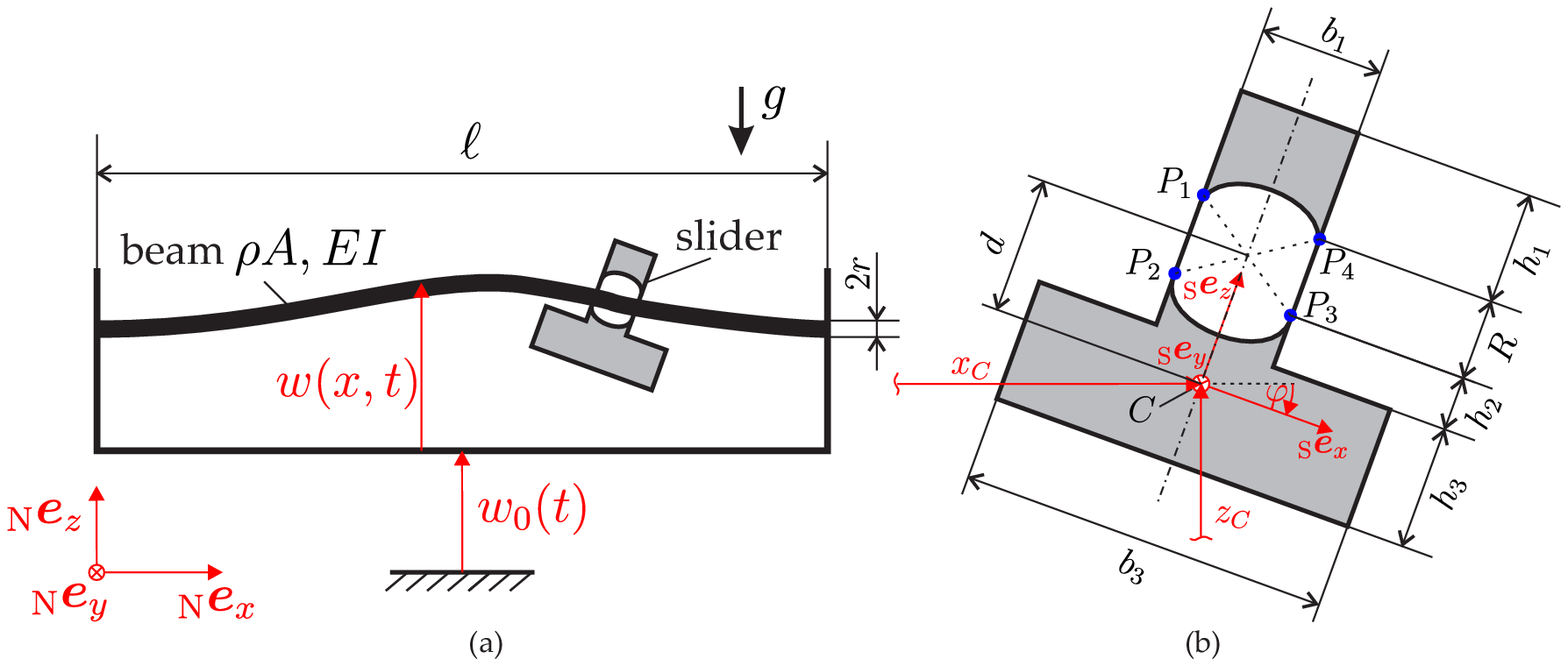}{Self-adaptive system: (a) base-excited beam with slider, (b) geometry of the slider}

\subsection{Component models\label{sec:decoupled}}
The considered system consists of two components, the beam and the slider. The beam is assumed to be slender, \ie, to possess a high length-to-thickness ratio, so that the Euler-Bernoulli theory is appropriate to describe its small linear-elastic bending deformations. The beam is assumed to be of a homogeneous material with density $\rho$, length $\ell$, constant cross section area $A$ and constant bending stiffness $EI$. The total deflection, measured from the inertial frame of reference (N), can be split into the base excitation $w_0(t)$ and the beam's elastic lateral deformation $w(x,t)$, see \fref{system_and_slider}a. The elastic deformation $w(x,t)$ of the beam is governed by the following partial differential equation and boundary conditions,
\ea{\rho A\left[ \ddot w_0(t) + \ddot w(x,t)\right] + EI w^{\prime\prime\prime\prime}(x,t) &=& 0 \quad x\in ]0,\ell[\fk \label{eq:pde}\\
w(0,t)=w^\prime(0,t)=w(\ell,t)=w^\prime(\ell,t) &=& 0\fp\label{eq:bcs}
}{do_not_reference_me}
Herein, prime and overdot indicate differentiation with respect to location $x$ and time $t$, respectively. Owing to the definition of $w(x,t)$ relative to the base excitation, the boundary conditions \erefo{bcs} correspond to a clamped-clamped configuration\footnote{A detailed derivation of \erefs{pde}-\erefo{bcs} from the standard formulation of the clamped-clamped Bernoulli beam is presented in \aref{beam_eqm_derivation}.}. The deformation due to its own weight is neglected\footnote{Alternatively, it could be assumed that the beam deforms to a straight geometry under gravitational acceleration.}.\\
The dynamical behavior of the elastic beam is approximated in terms of its leading $N_{\mathrm{mod}}$ natural modes of vibration $s_n(x)$ with associated modal coordinates $q_{\mathrm B,n}(t)$,
\e{w(x,t) \approx \tilde w(x,t) = \sum\limits_{n=1}^{N_{\mathrm{mod}}} s_n(x) q_{\mathrm B,n}(t) = \mms s\tra(x)\mms q_{\mathrm B}(t)\fp}{beam_discretization}
As indicated, the mode shapes are assembled in the vector $\mms s(x) = \matrix{ccc}{s_1(x) & \cdots & s_{N_{\mathrm{mod}}}(x)}\tra$ and the modal coordinates in the vector $\mms q_{\mathrm B}(t) = \matrix{ccc}{q_{\mathrm B,1}(t) & \cdots & q_{\mathrm B,N_{\mathrm{mod}}}(t)}\tra$. The mode shapes satisfy the common orthogonality and normalization conditions, namely, $\int_{0}^{\ell}\rho A \mms s(x)~\mms s\tra(x)\dd x = \mms I$, $\int_{0}^{\ell}EI \mms s^{\prime\prime}(x)~\left(\mms s^{\prime\prime}(x)\right)\tra\dd x = \operatorname{diag}\lbrace\omega_{\mathrm B,n}^2\rbrace$, where $\mms I $ is the identity matrix and $\omega_{\mathrm B,n}$ the natural frequency of the $n$-th mode.
\\
The slider considered in \cite{Miller.2013} occupies about one sixth of the beam length, so the dimensions of the slider are not significantly smaller than the dimensions of the beam itself. Therefore, opposed to the model suggested in \cite{Miller.2013}, we propose to model the slider as an extended body rather than as a point mass. This also permits us to consider multiple contact points of the slider on each side of the beam, so that an effective moment can be transmitted between the slider and the beam, instead of just a point load as in \cite{Miller.2013}. Considering the comparatively thin and thus much smaller stiffness of the beam, it seems appropriate to model the slider as a rigid body. Hence, the slider has three degrees of freedom, namely the horizontal translation $x_C$, the vertical translation $z_C$, and the rotation $\phi$ around the $y$-axis, see \fref{system_and_slider}b. These coordinates are assembled in the vector $\mms q_{\mathrm S} = \matrix{ccc}{x_C & z_C & \phi}\tra$. In the absence of contact interactions, the ordinary differential equations governing the slider's motion $\mms q_{\mathrm S}(t)$ read
\ea{\mms M_{\mathrm S}\ddot{\mms q}_{\mathrm S}(t) &=& -mg\mms e_z\fp}{eqm_slider}
Herein, $\mms M_{\mathrm S} = \operatorname{diag}\lbrace m,m,J_{yy}^{(C)}\rbrace$ is the slider's mass matrix, with $m$ begin the slider mass and $J_{yy}^{(C)}$ its mass moment of inertia with respect to center of mass $C$. Moreover, and contrary to the thin and slender beam, we assume that the slider is subjected to gravitational acceleration with coefficient $g$. The weight points toward the negative $z$-direction, where $\mms e_z = \matrix{ccc}{0 & 1 & 0}\tra$. 

\subsection{Coupling by contact constraints\label{sec:contact}}
A key feature of the proposed model is the clearance between the slider and the beam. The dimension $R$ of the hole in the slider is allowed to be larger than the thickness $2r$ of the beam, $R\geq2r$, \cf \fref{system_and_slider}. A concave geometry of the slider hole is assumed, as illustrated in \fref{system_and_slider}b. It follows that the beam can only come into contact with the slider at its four corners labeled by $P_1$ through $P_4$. More specifically, contact is only considered between $P_1$, $P_4$ at the top surface of the beam, and $P_2$, $P_3$ at the bottom surface.
\\
Dry contact between metallic bodies is assumed. In the normal contact direction, we account for unilateral interactions with potential lift-off. The relationship between normal gap and normal contact force is described in terms of the Signorini law. In the tangential direction, frictional interactions are considered. The relationship between tangential relative velocity and tangential contact force is modeled by the Coulomb law. This allows for a clear distinction between sticking and sliding friction. The limiting friction force of the Coulomb law is the product of the normal contact force and the friction coefficient $\mu$. The same friction coefficient is considered for both sticking and sliding. These \myquote{hard} contact laws do not capture the residual compliance stemming, for example, from elastic deformations of surface roughness asperities. This approach seems appropriate, since we expect macroscopic contact interactions, for which these contact laws are commonly considered to be valid. Beside the contact forces, possible contact impulses are taken into account by Newton's impact law.\\
The above-mentioned contact laws are set-valued, which generally gives rise to nonsmooth dynamics, with potential discontinuities on the velocity level at the instances of impact. This requires special attention when the system behavior is to be computed using time integration in order to ensure accurate modeling of the times and intensities of the impacts (as this is a critical issue for the accurate modeling of the overall non-smooth system dynamics). One approach to address this issue is the regularization of the contact laws. This results in single-valued contact laws so that standard methods for the numerical integration can be applied. However, this is well-known to give rise to a high numerical stiffness of the resulting ordinary differential equations. This, in turn, is known to cause detrimental effects on the numerical stability of the time integration procedure, and typically necessitates small time steps. As an alternative to regularization, the set-valued character of the contact laws can be taken into account explicitly by resorting to computational techniques that were developed specifically for nonsmooth dynamical systems. In this case, the ordinary differential equations are replaced by so-called \textit{measure differential inclusions} of the form \cite{Acary.2008,Leine.2013},
%
\e{\mms M\dd u - \mms h\left(\mms q,\mms u,t\right)\dd t - \mms W\left(\mms q\right)\dd \mms P = \mm 0\fp}{measure_di}
Consider first the case $\dd \mms P=\mm 0$. In this case, \eref{measure_di} simplifies to $\mms M\dd u = \mms h\dd t$, which can be transformed to a system of conventional ordinary differential equations in terms of the vector of generalized coordinates $\mms q = \matrix{cc}{\mms q_{\mathrm B}\tra & \mms q_{\mathrm S}\tra}\tra$ and velocities $\mms u = \matrix{cc}{\mms u_{\mathrm B}\tra &  \mms u_{\mathrm S}\tra}\tra=\dot{\mms q}$. In this case the corresponding mass matrix of the system is $\mms M = \mms{\operatorname{bdiag}}\lbrace\mms I, \mms M_{\mathrm S}\rbrace$, where $\mms{\operatorname{bdiag}}$ denotes the block-diagonal operator. The force vector $\mms h$ comprises both of internal and external forces,
\e{\mms h = \vector{- \operatorname{diag}\lbrace 2D_{\mathrm B,n~}\omega_{\mathrm B,n}\rbrace \mms u_{\mathrm B} - \operatorname{diag}\lbrace \omega_{\mathrm B,n}^2\rbrace\mms q_{\mathrm B} \\ -mg\mms e_z}\fp}{vector_h}
Beside the gravity force and elastic forces of the beam, viscous damping forces with the modal damping ratios $D_{\mathrm B,n}$ are also considered. These are used to describe weak dissipative effects related to both material and aerodynamical damping.\\
In the presence of contact ($\dd \mms P\neq\mms 0$), \eref{measure_di} can no longer be transformed to an ordinary differential equation system, but assumes the form of the more general measure differential inclusion. $\mms P$ represents the contact percussions, which denote an integral measure of both contact forces and impulses. $\mms W$ accounts for the contact kinematics and denotes the generalized directions of contact forces and impulses. As such, it depends on the generalized displacement $\mms q$. Contact percussions and relative displacements at the contact points must satisfy the contact laws, and are both not \textit{a priori} known.

\subsection{Time integration\label{sec:time_integration}}
In this work, Moreau's time stepping scheme was applied for the time discretization of \eref{measure_di} and the computational dynamic analysis. The contact laws were enforced using an augmented Lagrangian approach in conjunction with the Gauss-Seidel relaxation method, which involves sequential applications of basic proximal point operations. We refer to the appendix for details on the specific formulation of the contact laws, the kinematics, and on the computational method adopted for this study. Based on a preliminary convergence study, the time step $\Delta t$ was specified in such a way that the highest natural frequency $\omega_{\mathrm{B},N_{\mathrm{mod}}}$ was sampled with five time steps per period; \ie, $\Delta t = 2\pi/\left(5\omega_{\mathrm{B},N_{\mathrm{mod}}}\right)$. The overall simulation method was implemented in Matlab and was used to obtain the simulation results presented throughout this work.

\subsection{Nominal parameters\label{sec:nominal_parameters}}
The proposed model possesses many parameters. Since one of the primary goals was to reproduce the recent experimental observations by Miller \etal, the parameters were largely selected in accordance with \cite{Miller.2013}. Hence, the parameters listed in the left column of \tref{nominal_parameters} were directly adopted from one of the cases presented in \cite{Miller.2013}, whereas, the remaining parameters were selected according to the following discussion. It should be noted that the kinetic properties of the rigid slider and the contact kinematics are uniquely defined by the parameters given in \tref{nominal_parameters}. The remaining geometric parameters $h_1$ through $h_3$ and $b_3$, as defined in \fref{system_and_slider}b, are merely required for illustration, and so they are not provided.
\tab[b]{rc|rc}{quantity & value & quantity & value\\ \hline
$E$     & $131~\mathrm{GPa}$                    & $R/(2r)$                & $1.25$\\
$\rho$  & $8,333~\mathrm{kg}/\mathrm{m}^3$      & $\mu$                   & $0.1$\\
$\ell$  & $6~\mathrm{cm}$                       & $\epsilon_{\mathrm n}$  & $0.5$\\
$r$     & $0.2~\mathrm{mm}$                     & $\epsilon_{\mathrm t}$  & $0$\\
$m_{\mathrm{B}}$        & $0.3~\mathrm{g}$      & $D_{\mathrm{B},n}$      & $1\%$\\
$m$     & $0.8~\mathrm{g}$                      & $w_0(t)$                & $\hat w_0\cos\left(2\pi\fex t\right)$\\
$b_1$   & $\ell/6$                              & $\hat w_0$              & $0.1~\mathrm{mm}$\\
$d$     & $0.0067\ell$                          & $N_{\mathrm{mod}}$      & $5$\\
$J_{yy}^{(C)}$          & $0.006m\ell^2$        &&\\
$g$     & $9.81~\mathrm{m}/\mathrm{s}^2$        &&
}{Nominal parameters}{nominal_parameters}
\\
As will be shown later, the clearance between the slider and the beam plays an important role in the nonlinear dynamics and the passive self-adaptive behavior of this system. This is defined by the ratio $R/(2r)$ between the vertical distance of the slider's contact points $R$ and the beam's thickness $2r$. Its influence is analyzed in \sref{parameter_influences}. It was noted by the authors of \cite{Miller.2013} that large frictional effects might impair the self-adaptive behavior. Therefore, an appropriate surface treatment (polishing, lubrication) was needed in their experimental work. Consequently, a comparatively small friction coefficient was specified in this work. The effect of this parameter is also presented in \sref{parameter_influences}. It is generally difficult to accurately determine an appropriate value for the normal coefficient of restitution $0\leq\epsilon_{\mathrm n}\leq 1.0$. Based on preliminary parameter studies, fortunately, this parameter was found to have only a comparatively small effect on the steady-state dynamical behavior, so a value of $0.5$ was used throughout this work. The tangential coefficient of restitution can be used to account for elastic deformations in the tangential direction and plays an important role for the modeling of certain materials such as rubber. In the present case, where metallic bodies are considered, these effects have been neglected ($\epsilon_{\mathrm t}=0$).\\
The joint effect of material damping and aerodynamical damping under ambient conditions was described in terms of modal damping of the beam's natural modes. To this end, equal modal damping ratios of $1\%$ were specified. Based on preliminary parameter studies, it was found that this parameter had the usual strong effect on the vibration level near the quasi-linear resonance peak. In the relevant operation regime, where the system exhibits self-adaptive behavior, however, the effect of modal damping was found to be comparatively small. Regarding the forcing term, only pure sinusoidal excitation was considered in this work. The excitation level has an important influence. In particular, the excitation level must be large enough to trigger the self-adaptive behavior, as shown in \sref{parameter_influences}. The nominal value was selected sufficiently large. Finally, the number of natural modes $N_{\mathrm{mod}}$ has been specified in accordance with a preliminary convergence study.
\fighw[t]{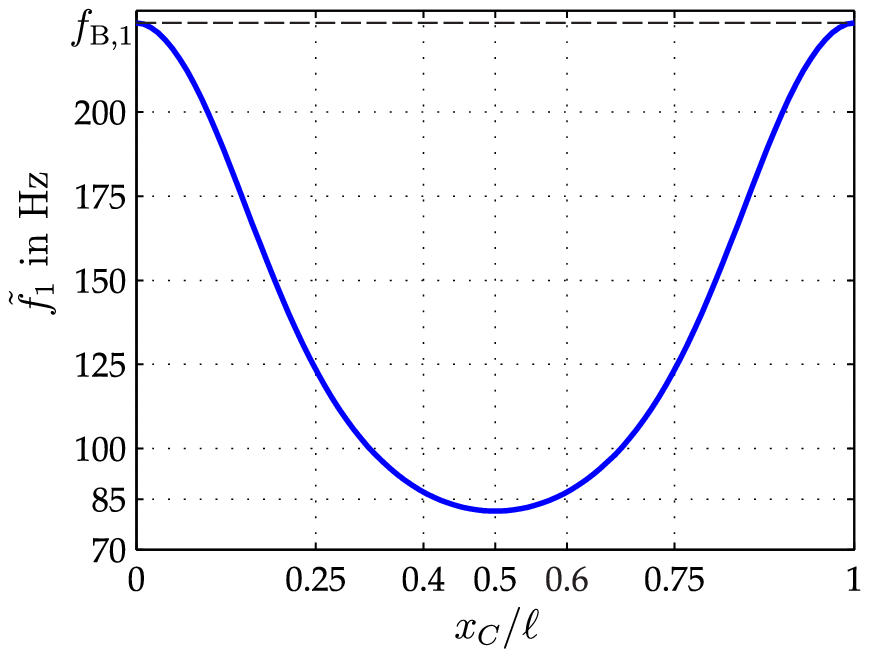}{Approximation of the system's first natural frequency for bonded slider as a function of the slider's axial location}
\\
The mass of the slider is $2.7$ times as large as the beam's mass. Thus, the slider has an essential influence on the system's natural frequencies, even if it is bonded to the beam at a certain location. This work focuses on the excitation frequency range in the neighborhood of the first system mode. The first mode's natural frequency $\omega_1$ for the beam with bonded slider can be approximated using the Rayleigh quotient. The first beam mode without the slider is considered as the ansatz function. Its modal deflection shape is $s_1(x)$ and its natural frequency is $\omega_{\mathrm B,1}$. When the slider is bonded at the location $x_C$, the Rayleigh approximation for the first modal frequency yields,
\e{\omega_1^2 \lesssim \frac{\omega_{\mathrm B,1}^2}{1+ms_1^2\left(x_C\right) + J_{yy}^{(C)}\left(s^\prime_1\left(x_C\right)\right)^2}=:\tilde\omega_1^2\fp}{rayleigh}
Herein, the second and third term in the denominator account for the influence of the mass and the mass moment of inertia of the slider, respectively. It can generally be stated that, for the parameters listed in \tref{nominal_parameters}, the rotation of the slider has a much smaller influence than the translation on the first natural frequency. The results are illustrated in \fref{natural_frequency_approx}. Frequencies $f$ and angular frequencies $\omega$ are related via $\omega = 2\pi f$. If the slider is bonded to the clamping ($x_C=0$ or $x_C=\ell$), it does not influence the system's natural frequency so that $\tilde f_1=f_{\mathrm B,1}$. The first mode has its only antinode at $x_C=0.5\ell$. At this location, the slider has the strongest influence. The approximation in \eref{rayleigh} does not account for the clearance between the slider and the beam. However, it can be useful to obtain a rough estimate of the relevant frequency range for the self-adaptive behavior.

\section{Typical dynamical behavior\label{sec:typical}}
In this section, the dynamical behavior is analyzed for two representative cases. More specifically, two different excitation frequencies are considered with identical initial conditions. The global and local dynamics of the adaptation phase as well as the steady state are analyzed in detail. This lays the foundation for the interpretation of the simulation results in the subsequent sections.\\
The two different excitation frequencies considered in this section are $85~\mathrm{Hz}$ and $125~\mathrm{Hz}$. The slider is initially located slightly right from the beam's center at $x_C(0)=0.6\ell$. The slider is placed so that the beam is vertically centered between its upper and lower contact points; \ie, $z_C(0)=w_0(0)+w\left(x_C,0\right)-d$. All other initial coordinates and velocities are zero. This definition of the initial values is also used in the subsequent investigations, where this is referred to as \textit{trivial initial conditions}. The temporal evolution of the slider's axial location $x_C(t)$ and the elastic displacement $w(\ell/2,t)$ of the beam's center are depicted in \fref{typical_behavior_overview}.\\
For $\fex=85~\mathrm{Hz}$, the slider moves toward the beam's center ($x_C=0.5\ell$) and assumes a fix point. The depicted time spans corresponds to $\mathcal O\left(10^3\right)$ excitation periods. Due to the finite line width, the oscillations on the short time scale can hardly be recognized in \fref{typical_behavior_overview}c-d. The axial motion takes place on a much longer time scale. The excitation frequency is close to the natural frequency of the beam with bonded slider at its center, \cf \fref{natural_frequency_approx}. The vibration increases largely monotonically during the transient phase and reaches a high level, which remains almost constant after the steady slider location is attained.\\
For $\fex=125~\mathrm{Hz}$, the slider moves toward a mean location of $\overline x_C\approx 0.75\ell$. As opposed to the previous case, a certain residual axial oscillation remains and a fixed point is never reached. The average vibration level increases widely monotonically, but an apparently chaotic modulation remains after the transient phase.
\figw[t]{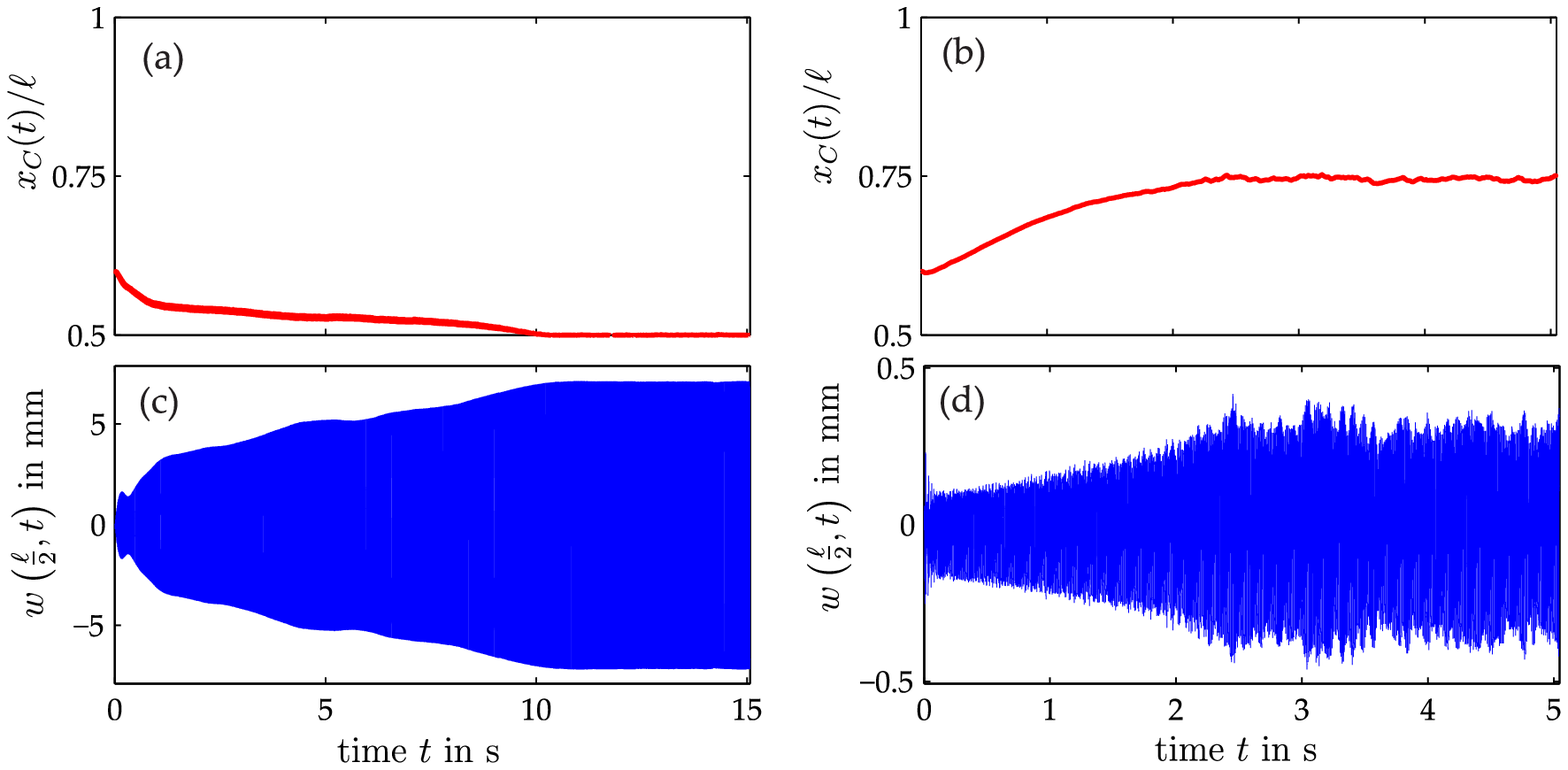}{Dynamical behavior for two different excitation frequencies: (a) and (c) $\fex=85~\mathrm{Hz}$, (b) and (d) $\fex=125~\mathrm{Hz}$; top: axial slider location, bottom: elastic displacement of the beam's center}
\\
Beside the different vibration levels and the dynamical nature (non-chaotic vs. chaotic) of the steady state, the two cases differ with regard to vibration form and contact pattern. The vibration form is illustrated in \fref{typical_behavior_detail} in terms of the modal energy distribution. Herein, the beam's modal energies $E_{\mathrm B,n}$ comprise both potential and kinetic terms, $E_{\mathrm B,n} = \frac12\omega_{\mathrm B,n}^2q_{\mathrm B,n}^2 + \frac12 u_{\mathrm B,n}^2$. The kinetic energy of the slider is split with regard to its coordinates into the translational components $E_{\mathrm{kin},x}=\frac12 m \dot x_C^2$ and $E_{\mathrm{kin},z}=\frac12 m \dot z_C^2$, and the rotational component $E_{\mathrm{kin},\phi}=\frac12 J_{yy}^{(C)} \dot\phi^2$. The plotting order ensured that no non-negligible energy component is hidden behind another.\\
As mentioned before, the slider assumes an axially centric location along the beam for $\fex=85~\mathrm{Hz}$. In this case, the slider undergoes pure lateral motion at steady state, while the beam vibrates in its first mode of vibration. Hence, the beam's vibration form is not significantly influenced by the slider. For $\fex=125~\mathrm{Hz}$, where the slider assumes an off-centric location along the beam, the slider's steady-state motion also exhibits a considerable rotational component. Moreover, the beam's motion is slightly distorted as higher modes also contribute to a certain extent.
\figw[t]{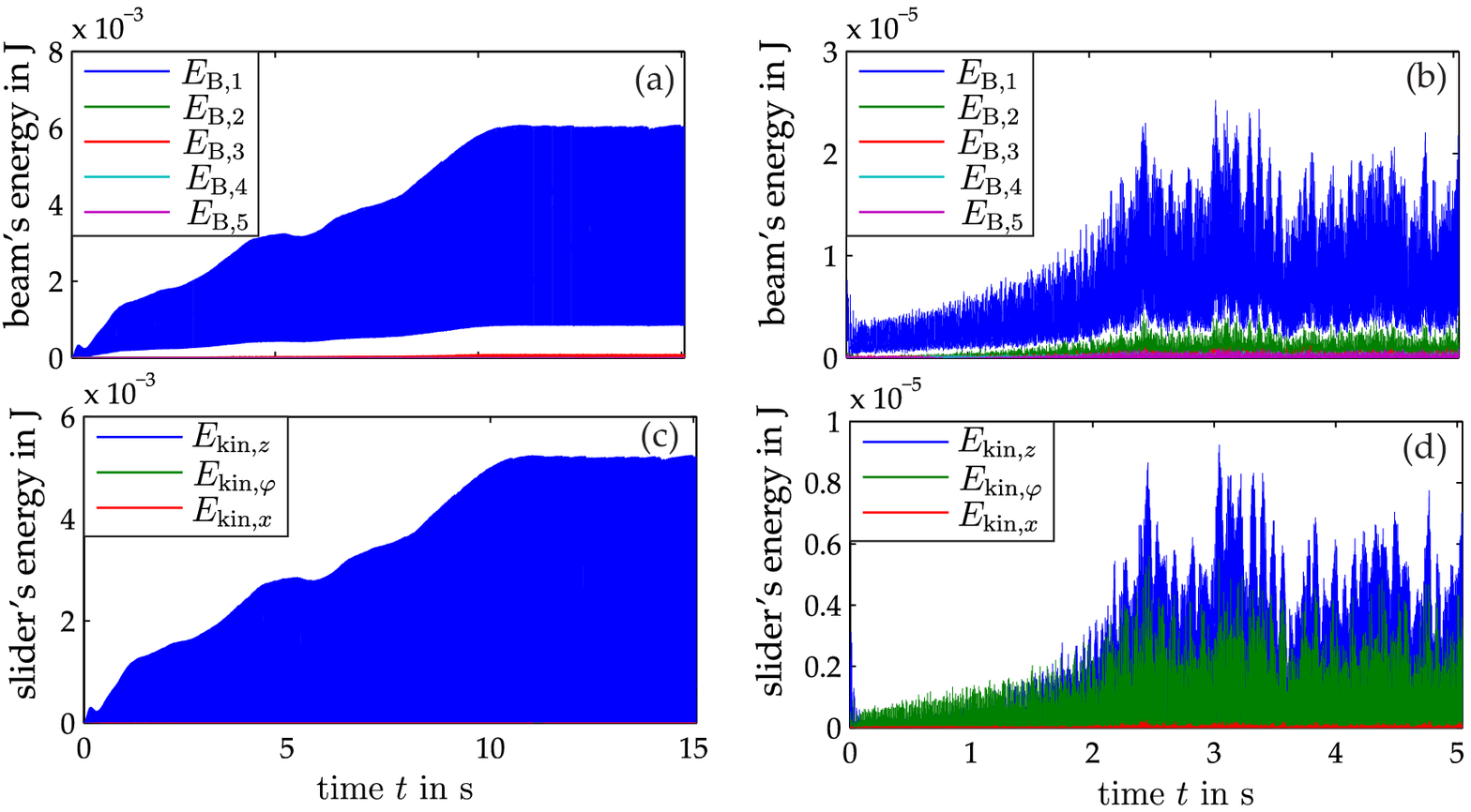}{Modal energy distribution for the two excitation frequencies considered in \fref{typical_behavior_detail}: (a) and (c) $\fex=85~\mathrm{Hz}$, (b) and (d) $\fex=125~\mathrm{Hz}$; top: mechanical energy in the beam modes, bottom: kinetic energy of the slider}
\\
Next, the contact pattern is analyzed. Only the respective steady regimes are considered. To this end, the time span is truncated to $12~\mathrm{s}\leq t\leq15~\mathrm{s}$ for $\fex=85~\mathrm{Hz}$ and to $3~\mathrm{s}\leq t\leq5~\mathrm{s}$ for $\fex=125~\mathrm{Hz}$. The excitation period was sampled by $128$ points and the corresponding contact states (separation, sticking, left-sliding, right-sliding) were determined. The results are illustrated in \fref{typical_behavior_contact}. Since the results are not (exactly) periodic, different contact states can be encountered at the sampling points of different excitation periods. At a certain time instant, the contact states are depicted with respect to their relative occurrence, for each of the four contact points $P_1$ through $P_4$. The occurrence is indicated by the gray shading: black means the contact state is attained every time, white means this state is never attained at the corresponding instant at the normalized excitation period.\\
As the symmetry of the system for $x_C=0.5\ell$ suggests, the contact pattern exhibits symmetry with respect to the vertical axis for $\fex=85~\mathrm{Hz}$, see \fref{typical_behavior_contact}a. This means that $P_1$ and $P_4$ ($P_2$ and $P_3$) are simultaneously separated or sticking, and slide simultaneously in opposite directions. Contact occurs either at the top ($P_1$ and $P_4$) or at the bottom ($P_2$ and $P_3$) of the beam. The separation phases overlap slightly, which indicates that there is a short time span where the slider is not in contact with the beam. During the downward motion of the beam, the slider separates from the bottom surface and impacts at the top surface of the beam. Until reaching its lower turning point, the top surface contracts, resulting in an outward-slip motion of the slider. When the beam starts moving upwards again, its top surface expands, leading to an inward-slip motion of the slider. The behavior is analogous in the neighborhood of the upper turning point of the beam. When the rotational motion of the slider and the effect of gravitational acceleration are neglected, the system is symmetric also with respect to the horizontal axis. The apparent symmetry of the response suggests that these effects play only an inferior role in this case.
\figw[t]{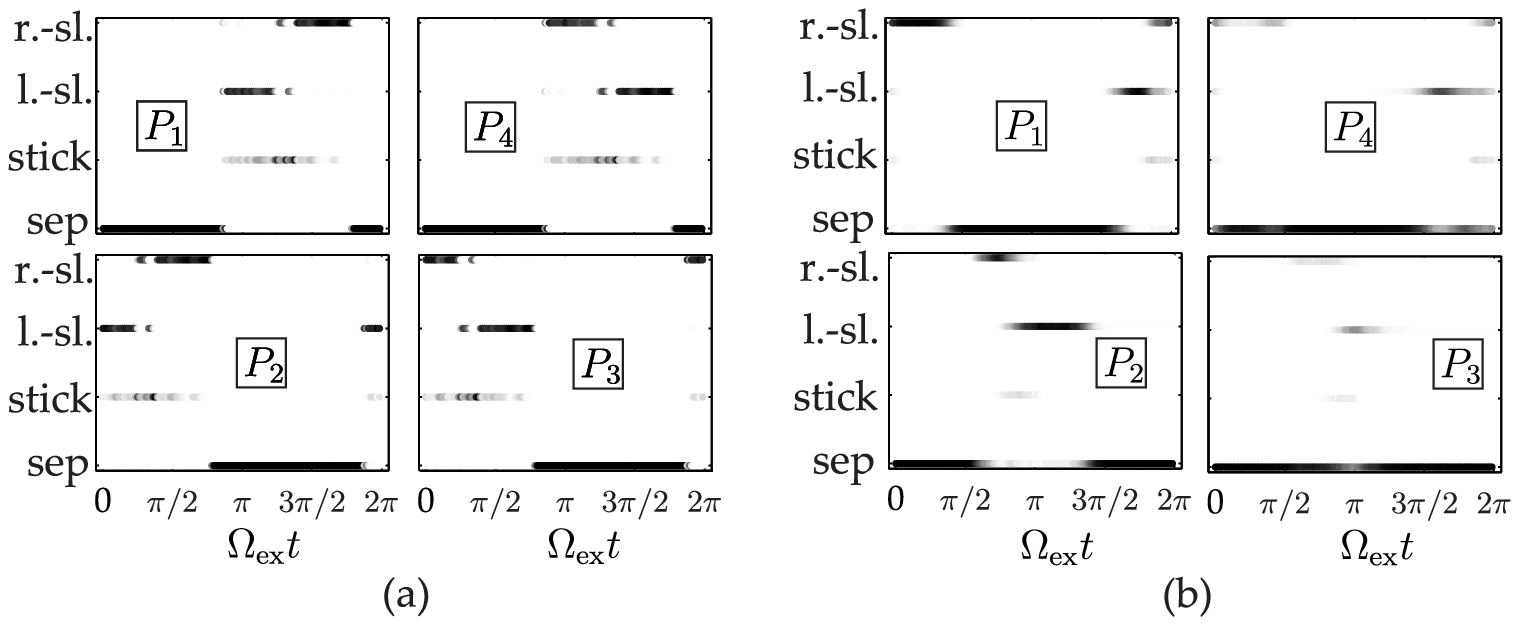}{Steady-state contact pattern for the two excitation frequencies considered in \fref{typical_behavior_detail}: (a) $\fex=85~\mathrm{Hz}$, (b) $\fex=125~\mathrm{Hz}$; the shading of the dots corresponds to occurrence over a sufficiently long steady-state time span}
\\
For $\fex=125~\mathrm{Hz}$, symmetry with respect to the vertical axis is not expected for the off-centric slider location $x_C\approx 0.75\ell$. Moreover, symmetry with respect to the horizontal axis is not evident. The reason for this is the complex interplay between rotational and translational motion. 
The energy between beam and slider is mainly transmitted through the points $P_1$ and $P_2$, whereas point $P_3$ is almost never in contact.

\section{Important parameter influences\label{sec:parameter_influences}}
In this section, the influence of different parameters on the self-adaptive behavior of the system is analyzed. In particular, we consider the case $\fex=125~\mathrm{Hz}$ and use the same initial values as specified in \sref{typical}. The dynamical behavior is simulated for different parameter values of (a) the excitation level $\hat w_0$, (b) the friction coefficient $\mu$, and (c) the clearance ratio $R/(2r)$. The resulting time evolution subject of the axial slider coordinate $x_C(t)$ subject to these changes is illustrated in \fref{th_xsl_125_pvar}.\\
If the excitation level is too small, the slider remains at its initial location. Beyond a certain threshold, the slider drifts to another mean location, which due to the system's nonlinearity, depends on the excitation level. This aspect will be further explored in \sref{unsteady}, where a periodic modulation of the excitation level is considered.\\
The friction coefficient also affects the steady-state mean location of the slider. In the special case of frictionless contact ($\mu=0$), the slider oscillates in the axial direction about the beam's center. This oscillation takes place on a much longer time scale than the forced vibrations. This corresponds to a completely different dynamical behavior. In this case, the system does not exhibit an off-centric steady-state (mean) location of the slider. Consequently, friction is necessary to explain the self-adaptive behavior.
\figw[t]{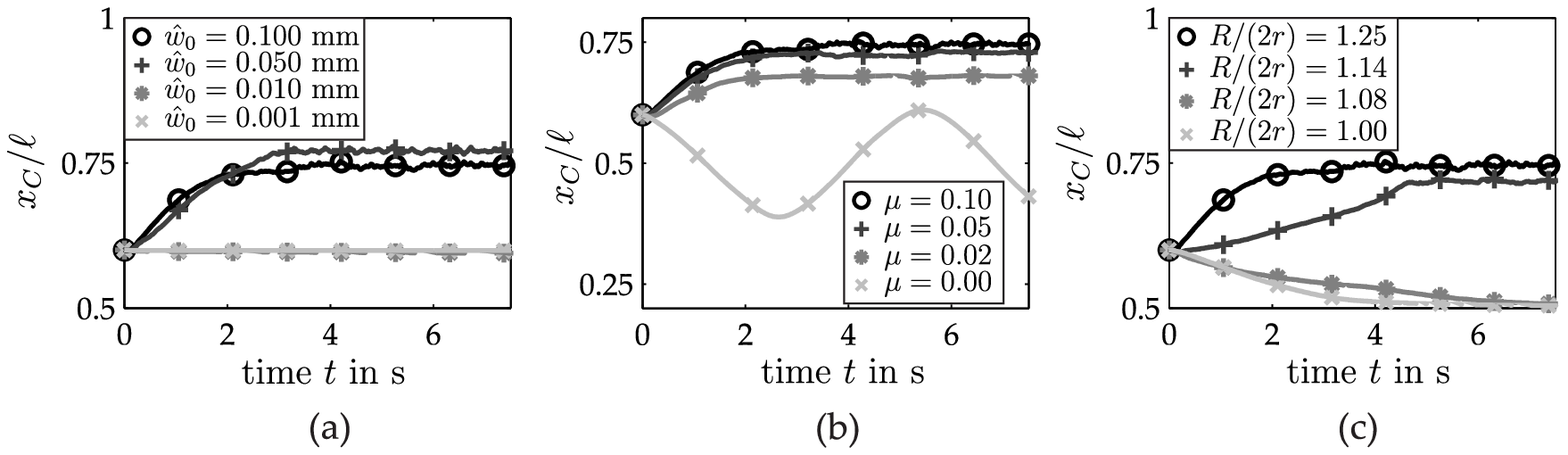}{Influence of certain parameters on the self-adaptivity for $\fex=125~\mathrm{Hz}$: (a) Effect of excitation level, (b) effect of friction, (c) effect of clearance}
\\
In \fref{th_xsl_125_pvar}c, the influence of the clearance ratio $R/(2r)$ is shown. Remarkably, the slider moves toward the beam's center, if the clearance is too small, and remains there, as the steady state is reached. If the clearance is sufficiently large, however, the slider can reach a steady-state mean location other than the beam's center. This special feature is essential for self-resonant behavior. We also note that the precise steady-state location of the slider is also influenced by the clearance.

\section{Steady-state operating range\label{sec:steady}}
In this section, the frequency-dependence of the self-adaptive behavior is analyzed in detail. If not otherwise noted, we limit the discussion to steady-state quantities, indicated by the superscript ($\mathrm{ss}$). In the case of non-periodic steady-state behavior the quantities are averaged over a sufficiently long time span, indicated by an overbar. In \ssref{operating_characteristic}, the operating range and its limits are presented. In \ssref{bifurcation}, a particular bifurcation is analyzed in order to highlight the potential dynamical richness of the system behavior.

\subsection{Frequency-dependence and limits of the operating regime\label{sec:operating_characteristic}}
The simulations were carried out for a wide range of excitation frequencies. Additionally, the initial axial slider location was varied in the range $0.5\ell\leq x_C(0)\leq \ell-b_1/2$. The upper limit $\ell-b_1/2$ ensures physical plausibility in the sense that the outer contact points of the slider do not reach beyond the clamped boundary. Otherwise, trivial initial conditions were specified. The resulting steady-state slider location and system energy are depicted in \fref{operating_characteristic_overview} a and b. For a specific excitation frequency, multiple steady-state (average) axial slider locations $\overline x_C^{\mathrm{ss}}$ and corresponding vibration behaviors may coexist, a feature that is typical of strongly nonlinear dynamical responses such as the ones encountered herein. The particular steady state location $\overline x_C^{\mathrm{ss}}$ reached depends on the initial value $x_C(0)$, since each of the co-existing solutions has its own domain of attraction. In \fref{operating_characteristic_overview}, the occurrence is again illustrated by gray shading: A black dot means that this steady state is reached regardless of the initial slider location, whereas, a white dot indicates that this steady state is never reached. In \fref{operating_characteristic_overview}b, the frequency response is also depicted for a non-adaptive reference, which is identical to the self-adaptive system depicted in \fref{system_and_slider} except that the axial location of the slider's center of mass is fixed to $x_C=\ell/2$. Therefore, the slider maintains its rotational and vertical translational degrees of freedom, but is not allowed to slide in the axial direction. Hence, the reference system undergoes contact interactions and, thus, exhibits nonlinear behavior, similar to the actual system with self-adaptiveness.
\figs[h!]{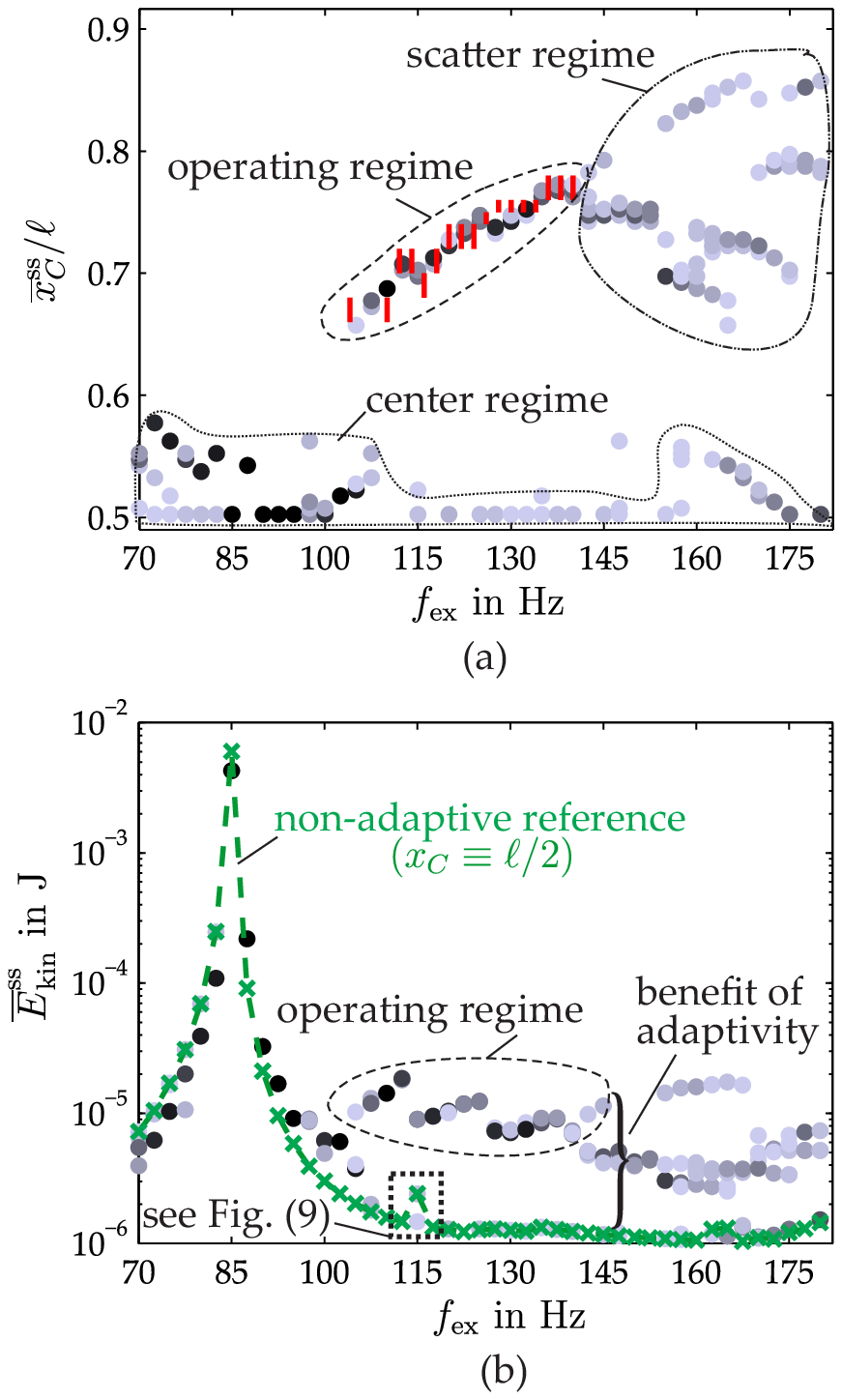}{Dependence of the steady-state behavior on the excitation frequency: (a) Axial slider location, (b) mechanical energy of the system; results were obtained for initial slider locations varied in the range $0.5\ell\leq x_C(0)\leq \ell-b_1/2$; the shading of the dots corresponds to their occurrence, as explained in the text}{0.85}
\\
The different steady states can be roughly grouped into operating, center and scatter regimes. For low ($\fex<105~\mathrm{Hz}$) and high ($\fex>175~\mathrm{Hz}$) excitation frequencies, the slider is likely to get stuck at (or very close to) the beam's center. The case $\fex=85~\mathrm{Hz}$ presented in \sref{typical} belongs to this \textit{center regime}. Beyond a certain frequency, $\fex\geq110~\mathrm{Hz}$, the \textit{operating regime} begins, where the slider is most likely to assume a mean axial position different than the center. Within this regime, $\overline x_C^{\mathrm{ss}}$ exhibits a characteristic dependence on the excitation frequency. This is required to achieve self-resonant behavior. Beyond an excitation frequency of $\fex\approx140~\mathrm{Hz}$, several distinct steady states $\overline x_C^{\mathrm{ss}}$ outside the beam's center coexist. This region is referred to as the \textit{scatter regime}. Due to the strong dependence on the initial slider location, no stable operation is expected in this regime, and the motion of the slider is chaotic. Thus, the center regime and the scatter regime limit the operating regime with respect to the excitation frequency.
The operating limits can also be inferred from \fref{basins_of_attraction}, where the basins of attraction for the different steady states are depicted. Within the operating range $110~\mathrm{Hz}\leq\fex\leq 140~\mathrm{Hz}$, the remaining center regime is much smaller than beyond these limits.
%
\figw[t]{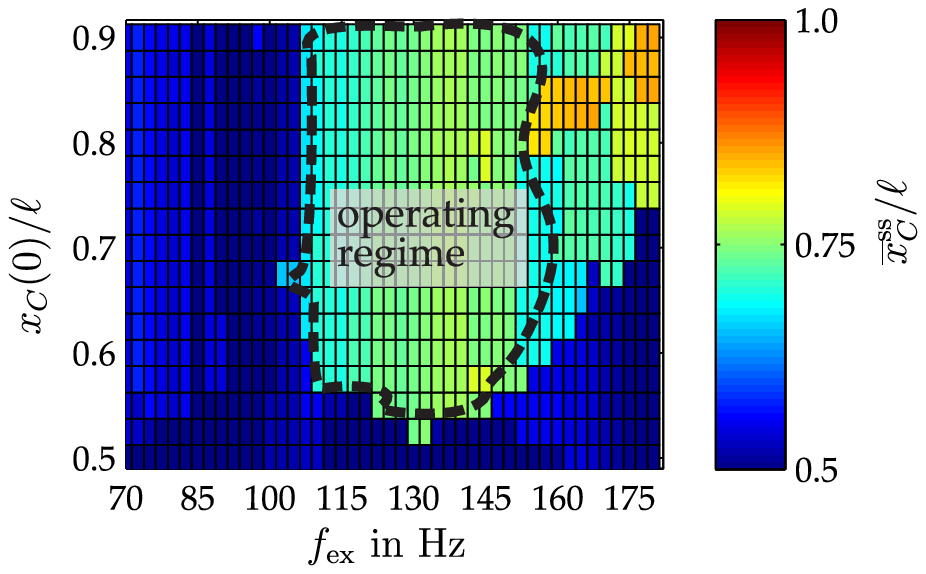}{Dependence of the steady-state axial slider location on its initial value ($t=0$), as a function of the excitation frequency}
\\
As can be seen in \fref{operating_characteristic_overview}b, the non-adaptive reference exhibits quasi-linear behavior in the low frequency range, with a resonance peak around $85~\mathrm{Hz}$. This resonance frequency is consistent with the approximate first natural frequency, \cf \fref{natural_frequency_approx}, and the resonance peak is located within the center regime. Therefore, the self-adaptive system behaves like the non-adaptive reference system and exhibits very similar performance in the neighborhood of resonance. In the comparatively wide operating range $110~\mathrm{Hz}\leq\fex\leq 140~\mathrm{Hz}$, the response level is fairly constant. For the considered set of parameters, the response level in this range is about two orders of magnitude lower than the quasi-linear resonance peak. Compared to the non-adaptive reference, the adaptive system extracts still considerably more energy from the excitation source in this regime. The benefit over the non-adaptive reference is about one order of magnitude here.
\\
The frequency dependence presented in \fref{operating_characteristic_overview} suggests a number of smaller or larger discontinuities. This is a characteristic property of systems with distinct states (stick, slip, lift-off of discrete contact points). The reason for this is that the contact pattern, in time and in space, may change instantaneously from one frequency to another. Due to the nonlinearity, moreover, different vibration states may coexist for the same combination $\lbrace\fex,x_C(0)\rbrace$. Different initial values for the remaining coordinates and velocities may thus lead to different steady states. Finally, the strongly nonlinear character of the contact interactions may trigger rich dynamical behavior including chaotic, sub- and superharmonic resonances. In the following subsection, such a discontinuity due to bifurcation is exemplified.

\subsection{Analysis of a particular bifurcation\label{sec:bifurcation}}
\figw[b]{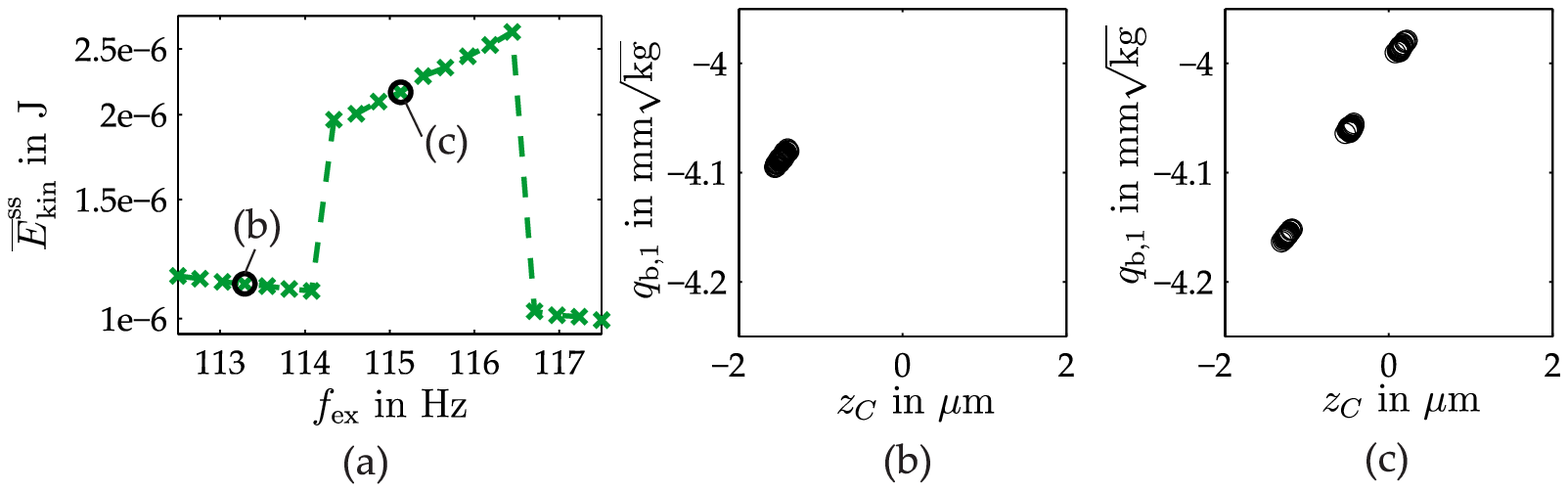}{A bifurcation of the non-adaptive reference system: (a) zoom into \fref{operating_characteristic_overview}b, (b) and (c) Poincar\'{e} maps for the points indicated in (a)}
Here, one specific discontinuity in the forced response is analyzed, namely that of the reference system, as indicated in \fref{operating_characteristic_overview}b.
In \fref{larry}a, a zoom into the frequency response in this range is presented. Apparently, the considered regime consists of two distinct branches. Note that for each solution point, trivial initial conditions were specified. Hence, only one solution point is obtained for each excitation frequency, and no information can be inferred regarding the stability of the continuations of these branches beyond the depicted frequency range.\\
To visualize the qualitative dynamical behavior on the two branches, Poincar\'{e} maps were derived for the points on either branch indicated in \fref{larry}a. To this end, the state is projected onto the $q_{\mathrm B,1}$-$z_C$ plane for time instants $t_n=n T_{\mathrm{ex}}$, where $T_{\mathrm{ex}}= 1/f_{\mathrm{ex}}$ is the excitation period and $n$ is a positive integer. The results are shown only for the steady regime in \fref{larry}b-c. In both cases, the Poincar\'{e} map consists of a (certain number of) agglomeration points. This indicates a nearly periodic behavior, with a narrow-bounded chaotic modulation. The essential difference between \frefs{larry}b and c is the number of agglomeration points. Apparently, a period tripling bifurcation takes place and the upper branch corresponds to a nearly period-three vibration.

\section{The notion of resonance\label{sec:resonance}}
As demonstrated in \sref{steady}, the system adapts itself in such a way that it outperforms the non-adaptive reference in terms of vibration response level. We would like to stress, however, that this alone does not justify refering to the system as being \myquote{self-resonant}. In order to attribute the capacity of self-resonance, one needs an appropriate definition of resonance for the adaptive system. The purpose of this section is to propose such a definition and to investigate if the system is indeed in resonance within its operating regime.\\
Consider the surrogate system illustrated in \fref{surrogate_system}. It differs from the adaptive system in \fref{system_and_slider} by the bearing which fixes the axial location of the slider's center of mass $x_C$. The design of the surrogate system is based on the observation that the slider assumes an (almost) constant axial location under steady-state conditions. In general, a reaction force $F_x$ is required to keep the slider at a certain location $x_C$. This reaction force balances the contact forces which would otherwise drive the slider to the left or to the right. Note that, since the bearing is attached to the slider's center of mass, the inertia effects do not contribute directly to the reaction force. In our study, for a fixed excitation frequency, the relationship $\overline F_x^{\mathrm{ss}}(x_C)$ is computed. As before, the initial energies are set to zero in the sense of the trivial initial conditions. The adaptation process takes place on a much longer time scale than the oscillations, so it can be regarded as quasi-steady. We therefore focus exclusively on the steady state dynamical behavior. A positive value $\overline F_x^{\mathrm{ss}}>0$ corresponds to an effective contact force that would drive the slider to the right, whereas a negative value $\overline F_x^{\mathrm{ss}}<0$ corresponds to a similar effective force toward the left. A value of zero means that there is no effective force acting either towards the left or the right. It should be emphasized that a zero mean value $\overline F_x^{\mathrm{ss}}=0$ does not imply a vanishing time-dependent reaction force; \ie, $F_x^{\mathrm{ss}}(t)\neq 0$ in general. Hence, the axial bearing influences the dynamics of the surrogate system, even at points with $\overline F_x^{\mathrm{ss}}=0$.
\figw[t]{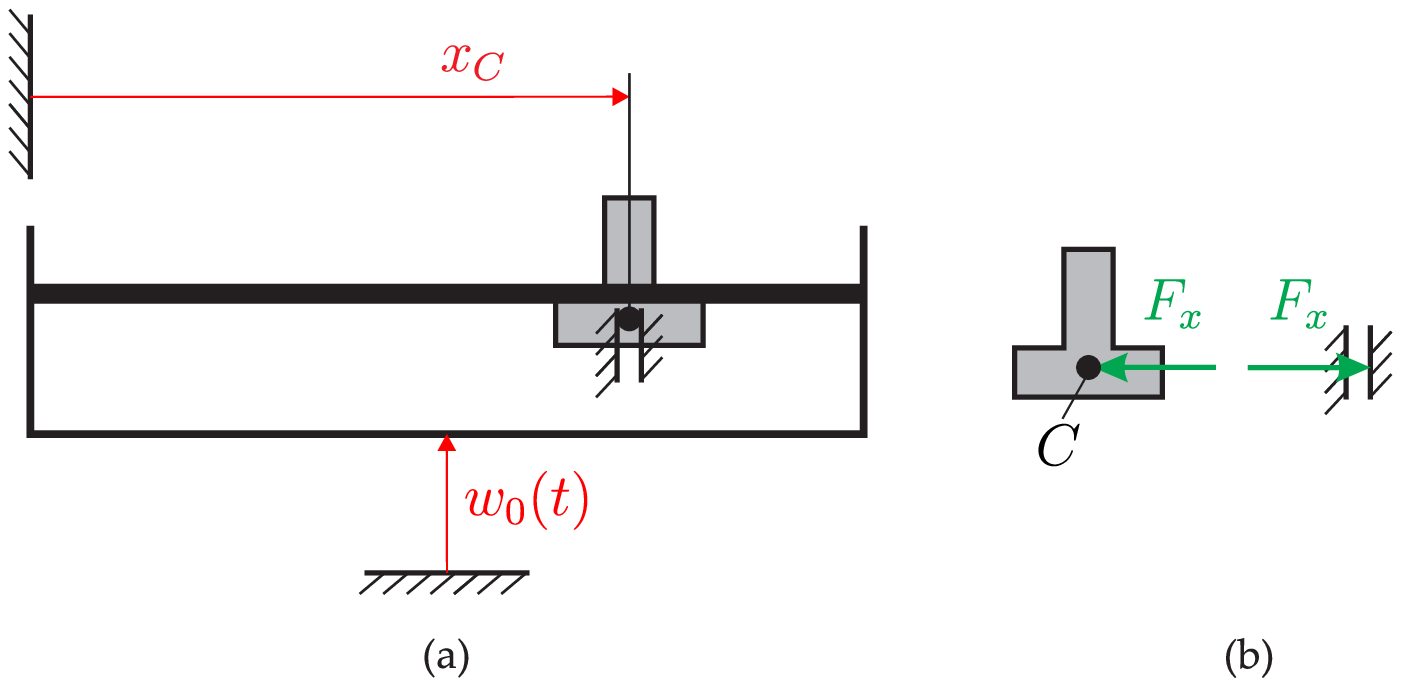}{Model of the non self-adaptive surrogate system: (a) beam with axially fixed slider, (b) definition of the reaction force $F_x$}
\\
Consider the relationship $F_x^{\mathrm{ss}}(x_C)$ illustrated in \fref{surrogate_frc_frfs}a for an excitation frequency of $130~\mathrm{Hz}$.
Distinct ranges of $x_C$ with positive or negative effective reaction forces exist, separated by zero crossings. At locations with $\overline F_x^{\mathrm{ss}}=0$ the slider might assume a fix point in the actual system. The fix point at $x_C\approx 0.75\ell$ is stable since a movement toward the left would result in a positive effective force would drive the slider back toward the right. On the other hand, a movement toward the right from this point would yield a negative effective force that would drive the slider back toward the left. It follows that the strength of the force is a measure of how intensely (and, thus, fast) the slider is driven along the axial direction. Consequently, coming from the center of the beam, the slider is expected to approach the point $x_C\approx 0.75\ell$ much more slowly than from the outside. This is consistent with numerical observations of the adaptive system.\\
In a small range $x_C\gtrsim0.5\ell$, it holds that $\overline F_x^{\mathrm{ss}}<0$, which indicates that the center of the beam is a stable fix point. However, slightly outside the center, an unstable fix point gives rise to a wide regime with positive axial force, which would drive the slider toward the stable fix point at $x_C\approx 0.75\ell$. The coexistence of fix points at the center and at a location $x_C\approx 0.75\ell$ is consistent with the operating characteristic illustrated in \fref{operating_characteristic_overview}a for the considered frequency $\fex=130~\mathrm{Hz}$. This suggests that $\overline F_x^{\mathrm{ss}}\left(x_C\right)=0$ provides a good estimate for $\overline x_C^{\mathrm{ss}}$ of the actual system, and that the surrogate and the adaptive systems exhibit very similar dynamical behavior at this location. This hypothesis is further evaluated in the following exposition by predicting the dependence $\overline x_C^{\mathrm{ss}}\left(\fex\right)$ using the surrogate system and comparing the results to the reference one.\\
To determine the operating characteristic, $\overline F_x^{\mathrm{ss}}$ was computed for a discrete set of frequencies in the range $105~\mathrm{Hz}\leq\fex\leq140~\mathrm{Hz}$, and a discrete set of axial locations in the range $0.6\ell\leq x_C\leq 0.8\ell$. For each frequency, the locations nearest to the zero crossing of $\overline F_x^{\mathrm{ss}}$ were determined and used as an estimate for $\overline x_C^{\mathrm{ss}}$ and illustrated as short red bars in \fref{operating_characteristic_overview}a. The surrogate-based prediction of the actual operating characteristic is quite good. This supports the aforementioned hypotheses and suggests that the surrogate system can be useful for the explanation of the adaptive system's behavior.
\figw[t]{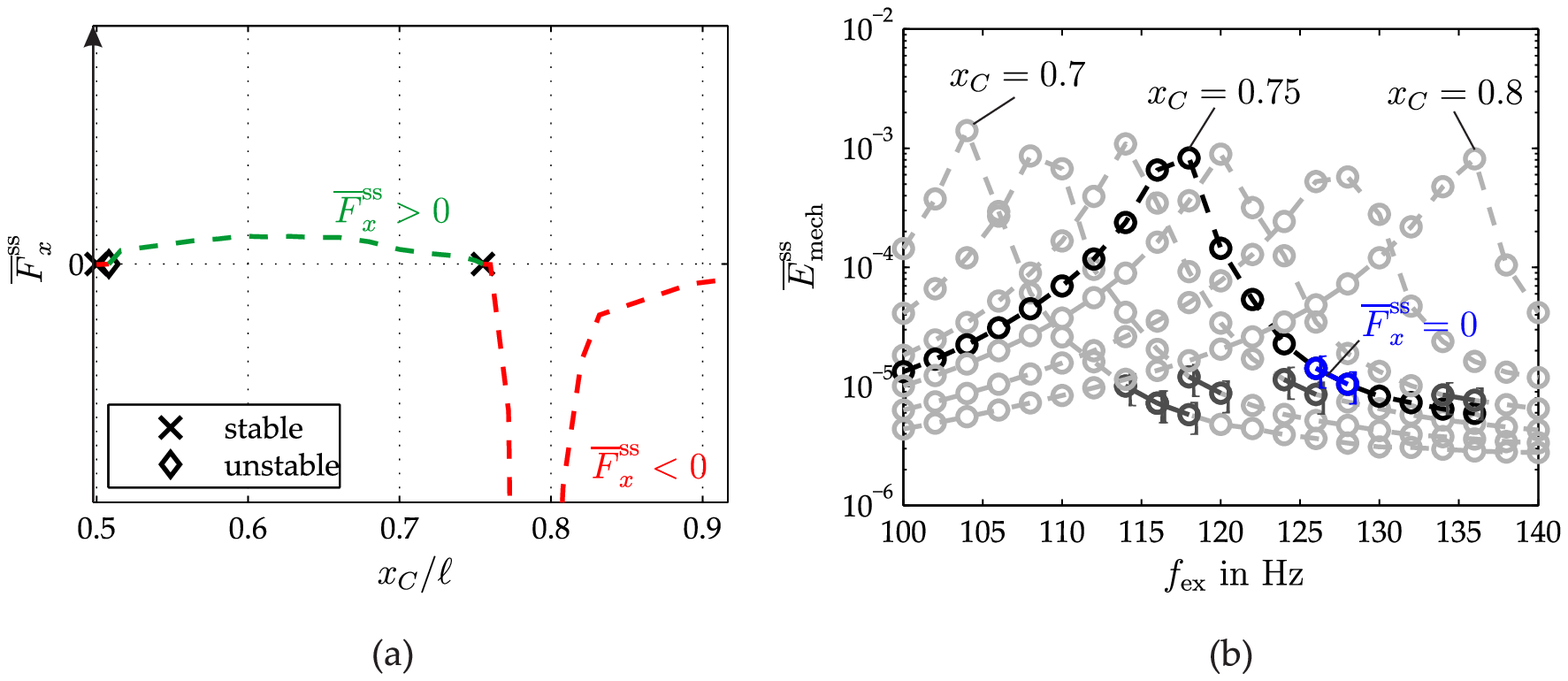}{Steady-state behavior of the surrogate system: (a) axial location dependence of the mean force and predicted fix points for $\fex=130~\mathrm{Hz}$, (b) fixed-slider frequency response in terms of the mechanical energy}
\fighw[t]{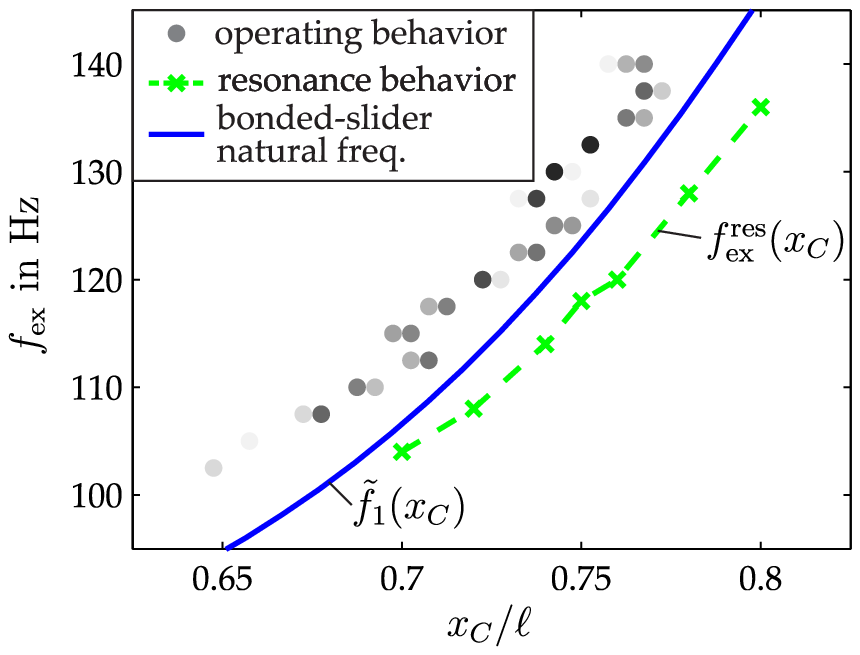}{Comparison of bonded-slider natural frequency, surrogate-based predicted resonance frequency and current operating behavior as a function of the slider coordinate}
\\
Motivated by the comparable responses of the surrogate and adaptive systems, we propose to define the notion of resonance on the basis of the surrogate system. Indeed, for a fixed slider location $x_C$, the steady-state frequency responses are determined and depicted in \fref{surrogate_frc_frfs}b. The ranges in which $\overline F_x^{\mathrm{ss}}$ crosses zero with respect to the frequency are highlighted and enclosed in square brackets. Apparently, the self-adaptive behavior is \textit{far from being resonant} for the considered set of parameters. For each frequency, it is possible to select a location $x_C^{\mathrm{res}}(\fex)$ that leads to a maximum vibration level. Conversely, it is possible to determine the resonance frequency $f_{\mathrm{ex}}^{\mathrm{res}}(x_C)$ as a function of the slider location. This predicted resonance frequency is illustrated in \fref{surrogate_fres}, and compared with the present operating behavior and the natural frequency approximation for the beam with bonded slider (\cf \fref{natural_frequency_approx}). Compared with the present operating behavior, this resonant location could provide an increase of the mechanical energy of about two orders of magnitude. Hence, the behavior of the system could be significantly improved by manually adjusting the slider location to $x_C^{\mathrm{res}}$ for each frequency $\fex$, rather than permitting the system to tend towards $\overline x_C^{\mathrm{ss}}\left(\fex\right)$ itself as in the present, self-adaptive configuration. This immediately gives rise to the question, of whether the self-adaptive system could be modified in such a way that its operating characteristic gets closer to the optimal case. In this context, the proposed method for estimating the resonant and the operating characteristics could be useful design tools.

\section{Unsteady operation\label{sec:unsteady}}
This study was so far limited to examining the steady-state response of the system subject to harmonic excitation with constant amplitude $\hat w_0$ and frequency $\fex$. In a real environment, however, these parameters may vary over time. It is the purpose of this section, to assess the performance of the self-adaptive system under unsteady operating conditions and to highlight certain resulting dynamical phenomena. In \ssref{hysteresis}, the influence of a time-dependent frequency is analyzed, and in \ssref{ampmod}, the case of a modulated excitation level is considered.

\subsection{Excitation frequency variation\label{sec:hysteresis}}
In this subsection, the dynamical behavior in the presence of a sine sweep excitation is studied. Two cases are considered, namely an up-sweep and a down-sweep, each in the range $70~\mathrm{Hz}\leq\fex\leq150~\mathrm{Hz}$. Thus, the lower frequency is far beyond the operating range. The slider is initially placed at the position $x_C=0.6\ell$, and otherwise trivial initial conditions are specified. The behavior is determined for both the self-adaptive system and the non-adaptive reference system (\ie, the system with the axially fixed slider at $x_C=0.5\ell$). The results are illustrated in \fref{sweeps}.
\figw[b]{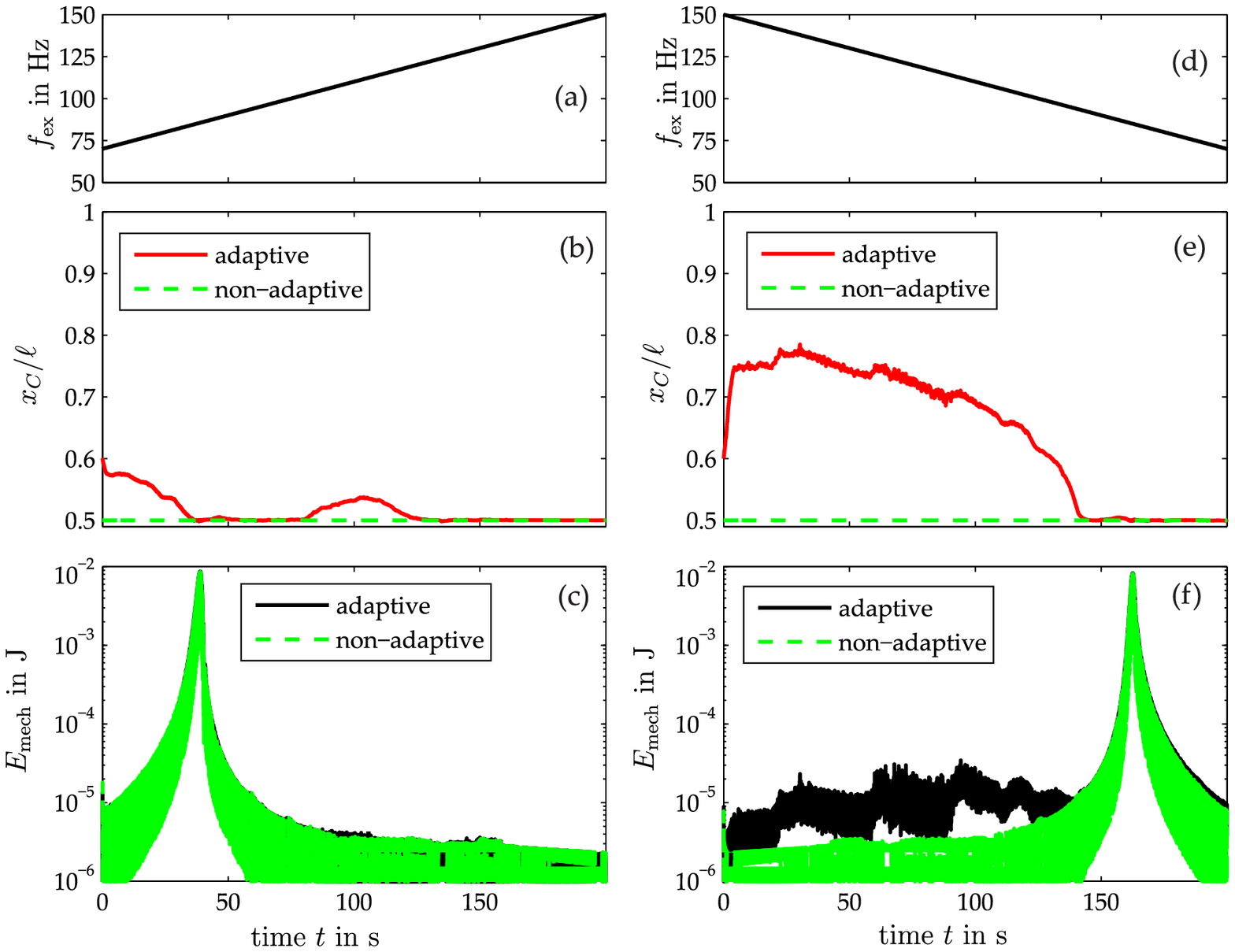}{Dynamical behavior for sine sweep excitation: (a)-(c) up-sweep, (d)-(f) down-sweep; top: excitation frequency, center: axial slider location, bottom: mechanical energy}
\\
In the up-sweep, the slider moves to the center of the beam and remains in the neighborhood of the center henceforth. Consequently, the adaptive system's response closely follows that of the non-adaptive system. In contrast, the slider attains the operating regime in the down-sweep. The resulting vibration level is considerably larger than that of the non-adaptive system. Before the primary resonance frequency is reached, the slider moves to the center of the beam and stays there. The results are consistent with the steady case, see \fref{operating_characteristic_overview}.\\
The results for the sine sweeps indicate that the slider can get stuck at the center of the beam under unsteady operating conditions. Hence, measures will be needed in practice to avoid the slider exceeding its intended operating regime. Otherwise, the coexistence of multiple stable steady-state slider locations can cause hysteresis effects with respect to a variation of the excitation frequency.

\subsection{Excitation level modulation\label{sec:ampmod}}
\figw[b!]{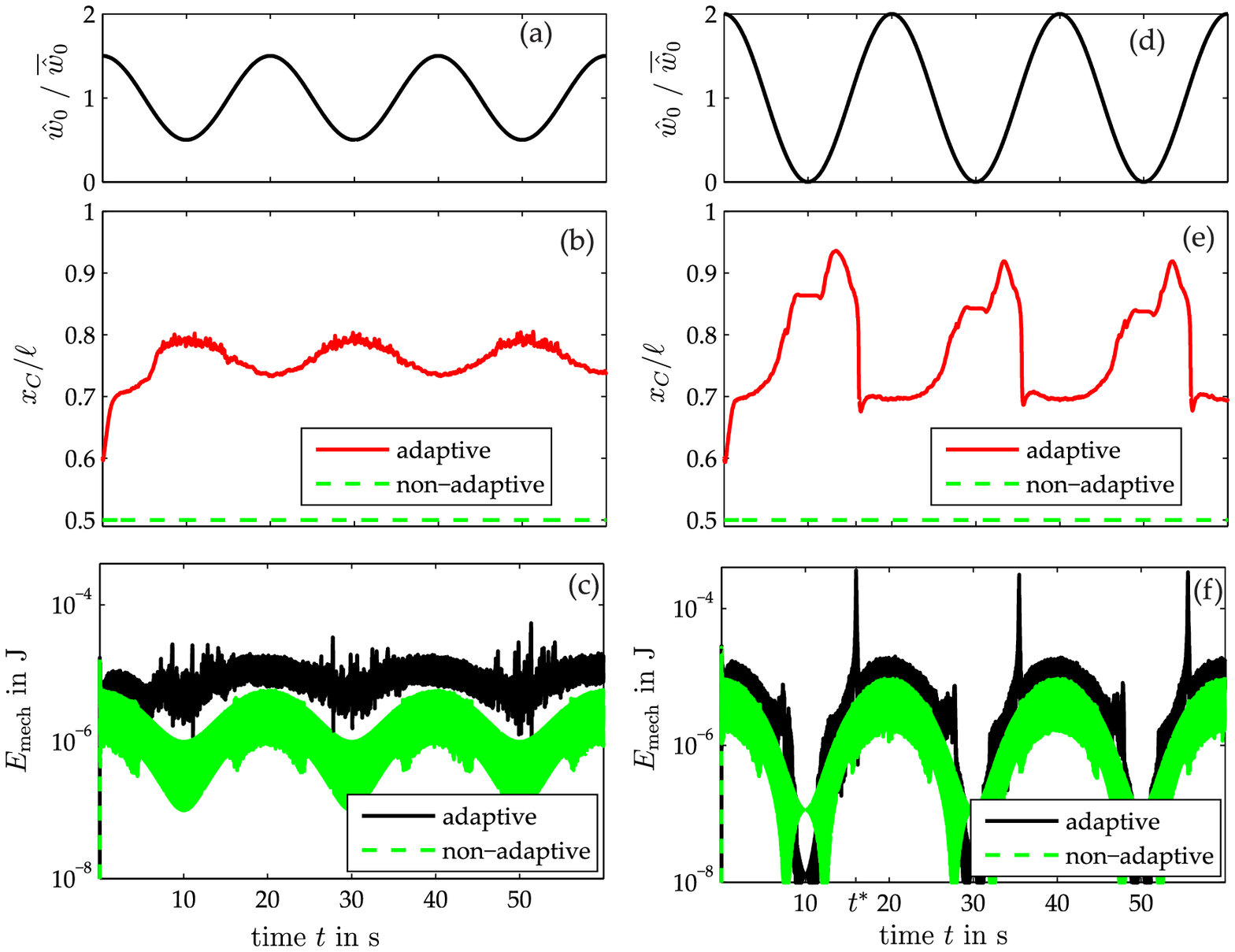}{Dynamical behavior for modulated excitation level at $\fex=135~\mathrm{Hz}$: (a)-(c) moderate modulation intensity $\alpha=0.5$, (d)-(f) strong modulation intensity $\alpha=1.0$; top: excitation level, center: axial slider location, bottom: mechanical energy}
In a second set of simulations, the excitation frequency was fixed to $\fex=135~\mathrm{Hz}$. Instead, the excitation level undergoes sinusoidal oscillation with $\hat w_0(t) = \overline{\hat w}_0\left[1 \right.$ $\left.+ \alpha\cos\left(2\pi f_{\mathrm{a}}t\right)\right]$, around the average value $\overline{\hat w}_0=0.1~\mathrm{mm}$, where $0\leq \alpha\leq 1$ controls the modulation intensity and $f_{\mathrm a}=0.05~\mathrm{Hz}$ denotes the modulation frequency. For the considered excitation frequency, there are thus $2,700$ excitation pseudo-periods during one modulation period. Again, the slider is initially placed at $x_C=0.6\ell$, and otherwise trivial initial conditions are specified. The dynamical behavior is determined for both the self-adaptive system and the non-adaptive reference system. The results are illustrated in \fref{ampmod} for two different modulation intensities, $\alpha=0.5$ and $\alpha=1.0$.
\figw[t!]{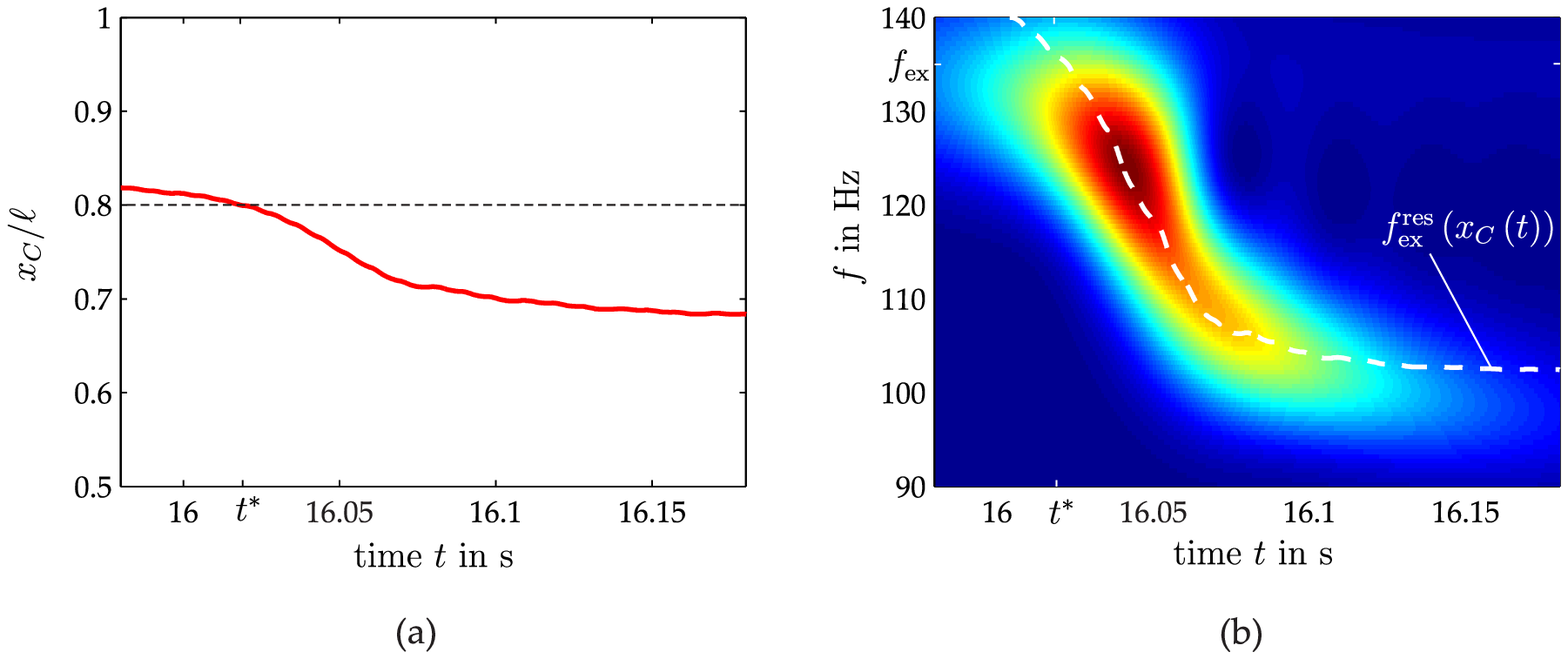}{Resonance capture encountered for modulated excitation level: (a) zoom into \fref{ampmod}e, (b) time-frequency dependence with regard to the beam's first modal coordinate $q_{\mathrm B,1}(t)$ where blue indicates a low and red a large magnitude}
\\
As noted already in \sref{parameter_influences}, the excitation level influences the quasi-steady axial slider location, so that $x_C$ oscillates almost periodically with the modulation frequency. In the case of the smaller modulation intensity, $x_C$ exhibits an erratic behavior on the short time scale. This effect appears magnified near $x_C=0.8\ell$. In this range, the mechanical energy shows certain \myquote{spikes}. For the larger modulation intensity, the oscillation of $x_C$ is more prominent. Moreover, larger spikes appear in the mechanical energy. The spike phenomenon is analyzed and explained in the following. It can generally be stated that the self-adaptive system outperforms the non-adaptive reference in terms of vibration response level. Apart from the occurrence of the spikes, it appears that the overall response level of the self-adaptive system is slightly less sensitive with respect to a variation of the excitation level.
\\
In \fref{ampmod_resonance_capture}a, a zoom is depicted into the time history of the slider location $x_C(t)$ in the neighborhood of the first large spike shown in \fref{ampmod}e. At time $t^*$, the slider reaches the location $x_C=0.8\ell$. In accordance with the investigations of the surrogate system, this is the resonant location for this excitation frequency, $x_C^{\mathrm{res}}\left(135~\mathrm{Hz}\right) = 0.8\ell$, or equivalently, $f^{\mathrm{res}}_{\mathrm{ex}}\left(0.8\ell\right)=135~\mathrm{Hz}$, \cf \fref{surrogate_frc_frfs}b and \fref{surrogate_fres}. The spikes in \fref{ampmod} are therefore interpreted as \textit{transient resonance captures} \cite{vaka2008b}. This can also be inferred from \fref{ampmod_resonance_capture}b, where the time-frequency dependence of forced vibrations is illustrated in terms of the wavelet transform of the beam's first modal coordinate $q_{\mathrm B,1}(t)$. Dark red indicates a large response level, while dark blue indicates a small response level. In addition, the location-dependent resonance frequency $f^{\mathrm{res}}_{\mathrm{ex}}(x_C(t))$ is depicted, which was evaluated at the location $x_C(t)$ obtained from numerical integration of the actual system. Apparently, the time-frequency content of the vibration response closely follows the estimated resonant characteristic. This supports the hypothesis that the encountered spike is actually a transient resonance phenomenon. The transient resonance frequency $f^{\mathrm{res}}_{\mathrm{ex}}(x_C)$ coincides with the excitation frequency $\fex$ at time $t^*$. This is slightly before the maximum response level is reached. It should be noted that a time delay between resonance coincidence and maximum response is typical for transient resonance phenomena.\\
The investigations of the excitation level modulations demonstrate that the self-adaptive system in its current design may undergo (transient) resonance phenomena, or transient resonance captures, in the operating regime. However, steady resonance phenomena, where the system attains large vibration levels for a longer period of time, were not encountered. This is consistent with the coincidence of (shorter) spikes and erratic oscillations of $x_C$ in \fref{ampmod}b-c, which also suggests the static instability of the resonant location for the present system design.

\section{Conclusions\label{sec:conclusions}}
The simulation results presented within this work reproduce the experimental observations reported in \cite{Miller.2013} in the following important respects:
\begin{enumerate}[label=(\alph*)]
\item the (nearly) time-constant steady-state axial slider location,
\item the qualitative frequency dependence of this location,
\item the increased vibration level compared with the non-adaptive reference system, and
\item the limitations of the operating regime.
\end{enumerate}
Although a stable resonance, in accordance with the notion introduced in \sref{resonance}, was not encountered in the operating range for the specific set of model parameters, considered in our work, the self-adaptive behavior and its beneficial effects have been recovered and confirmed in our computational study. More specifically, the simulation results suggest that the adaptive system could provide a moderately constant power output in a wide frequency range, which would be particularly attractive for energy harvesting applications. With regard to the limitation of the operating regime, it was also found in the simulation that the self-adaptive behavior occurs only if the excitation level exceeds a certain critical threshold, and the excitation frequency is within certain bounds. Based on the good qualitative agreement between experiment and simulation, it is concluded that the proposed model properly describes the essential physical interactions underlying the self-adaptive behavior of the system.\\
New insight into the intriguing self-adaptive behavior were derived from our theoretical investigations. It was found that the self-adaptivity relies on the clearance between the slider and the beam. This is responsible for dynamical contact interactions that permit the self-adaptiveness of the system by
re-arrangement of the slider along the beam and the existence of a stable, steady-state mean location of the slider away from the antinode at the beam's center. Self-adaptivity was not encountered for too small clearances or in the frictionless case, and, therefore, it is concluded that both backlash and friction play a central role in this phenomenon. In spite of these nonlinear phenomena, the resulting beam's motion was found to be still dominated by its first mode. However, the strong nonlinearity stemming from the contact interactions contributes to the overall rich dynamical behavior, including chaotic motions, transient resonance captures, hysteresis effects, and the coexistence of multiple steady vibration states.
\\
Future work could be directed toward a quantitative experimental validation of the proposed model. In this regard, it would be desirable to directly observe contact interactions and analyze the time-frequency content of the system's vibration. Furthermore, the simulation results suggest that the general setup has a considerable margin for improvement. Several parameters, such as the clearance, the excitation level and the friction coefficient were found to have a significant influence on the operating behavior. Here, the model could be useful for the development of more effective passively self-adaptive system designs.

\begin{acknowledgements}
The first author gratefully acknowledges the Daimler and Benz Foundation for their support of the postdoctoral scholarship (project 32-05/15).\\
The second author gratefully acknowledges the European Commission for its support of the Marie Curie program through the ITN ANTARES project (GA 606817).
\end{acknowledgements}

\section*{Compliance with Ethical Standards}
Funding: The work leading to this publication was funded by the Daimler and Benz Foundation, and the European Commission, see acknowledgements.\\
Conflict of Interest: The authors declare that they have no conflict of interest.

\begin{appendix}
\section{Derivation of the beam's equation of motion\label{asec:beam_eqm_derivation}}
In this appendix, the beam's equation of motion \erefs{pde}-\erefo{bcs} is derived from the formulation found in any standard textbook on vibrations of continua.
The dynamic deformation $v(x,t)$ of the beam in its autonomous configuration, in accordance with the theory specified in \sref{decoupled}, is governed by
\ea{\rho A\ddot v(x,t) + EI v^{\prime\prime\prime\prime}(x,t) &=& 0 \quad x\in ]0,\ell[\fk \label{eq:pde_auto}\\
v(0,t)=v^\prime(0,t)=v(\ell,t)=v^\prime(\ell,t) &=& 0\fp\label{eq:bcs_auto}
}{do_not_reference_me}
Introducing the base excitation $w_0(t)$ to the autonomous system, according to the considered connection between base and beam, can be modeled by adjusting the boundary conditions in \eref{bcs_auto} to,
\e{v(0,t)=w_0(t), v^\prime(0,t)=v^\prime(\ell,t) = 0, v(\ell,t)=w_0(t) \fp}{bcs_absolute}
Note that \eref{pde_auto} remains unchanged.\\
A coordinate transform is now introduced, 
\e{v(x,t) = w(x,t) + w_0(t)\fk}{coordinate_trafo_beam}
where the dynamic deformation $v(x,t)$ is split into the imposed base displacement $w_0(t)$ and an elastic deformation $w(x,t)$ measured from the time-dependent base location. To apply the transform, also the time and space derivatives of \eref{coordinate_trafo_beam} are needed,
\ea{\ddot v(x,t) &=& \ddot w(x,t) + \ddot w_0(t)\fk \label{eq:coordinate_trafo_derA}\\
v^\prime(x,t) &=& w^\prime(x,t) \fk \label{eq:coordinate_trafo_derB}\\
v^{\prime\prime\prime\prime}(x,t) &=& w^{\prime\prime\prime\prime}(x,t)\fp \label{eq:coordinate_trafo_derC}
}{do_not_reference_me}
Substituting \erefs{coordinate_trafo_beam}-\erefo{coordinate_trafo_derC} into \eref{pde_auto} and \eref{bcs_absolute} yields,
\ea{\rho A\left[\ddot w_0 + \ddot w(x,t)\right] + EI w^{\prime\prime\prime\prime}(x,t) &=& 0 \quad x\in ]0,\ell[\fk \nonumber\\
w(0,t)=w^\prime(0,t)=w(\ell,t)=w^\prime(\ell,t) &=& 0\fp\nonumber
}{do_not_reference_me}
This is identical to \erefs{pde}-\erefo{bcs}, already introduced in \sref{decoupled}, which were used for the investigations in the present work. Note that  \eref{pde_auto} with boundary conditions \erefo{bcs_absolute} is fully equivalent with \erefs{pde}-\erefo{bcs}. Hence, either formulation can be utilized for the simulation of the coupled problem. Simulation results are therefore expected to be identical, in practice possibly limited by finite numerical precision.

\section{Contact laws\label{asec:contact_laws}}
In this appendix, the contact laws are formulated, which are used to describe the dynamic contact interactions between beam and slider. For contact point $P_i$, the Signorini law relates the normal contact force $\lambda_{\mathrm{n},i}$ and the normal contact gap (relative displacement) $g_{\mathrm{n},i}$. The law and can be compactly written as
\e{0 \leq \lambda_{\mathrm{n},i} \perp g_{\mathrm{n},i} \geq 0\fp}{signorini}
Physically, this means that only compressive forces are allowed, and no penetration is permitted. Moreover, the symbol $\perp$ indicates the complementarity condition, \ie, that the force can admit nonzero values only if the contact is closed ($g_{\mathrm{n},i}=0$), and the force must be zero when the contact is open ($g_{\mathrm{n},i}>0$).\\
The one-dimensional Coulomb law relates the tangential gap velocity $\gamma_{t,i}$ and the tangential contact force $\lambda_{\mathrm{t},i}$,
\e{\begin{cases}
\text{sticking:} \quad \gamma_{\mathrm{t},i} = 0\quad \land & \quad \left|\lambda_{\mathrm{t},i}\right| < \mu \lambda_{\mathrm{n},i} \\
\text{slipping:} \quad  \gamma_{\mathrm{t},i} \neq 0\quad \land & \quad \lambda_{\mathrm{t},i} = -\mu \lambda_{\mathrm{n},i} \operatorname{sign}\left(\gamma_{\mathrm{t},i}\right)
\end{cases}\fp}{coulomb}
Thus, the limit friction force $\mu \lambda_{\mathrm{n},i}$ depends on the normal contact force $\lambda_{\mathrm{n},i}$. The negative sign in the slipping case ensures that the friction force points in the opposite direction of the relative sliding velocity.\\
In the case of instantaneous impulses, the Newton impact law is considered. Normal and tangential impulses are denoted $\Lambda_{\mathrm{n},i}$ and $\Lambda_{\mathrm{t},i}$, respectively. The associated coefficients of restitution are $\epsilon_\mathrm{n}$ and $\epsilon_\mathrm{t}$. Velocities before and after the impulse marked by the superscripts $+$ and $-$. Applied to the normal contact, this leads to
\e{0 \leq \Lambda_{\mathrm{n},i} \perp \gamma_{\mathrm{n},i}^+ + \epsilon_\mathrm{n}\gamma_{\mathrm{n},i}^- \geq 0\fk}{newton_normal}
while for the tangential contact, this results in
\e{\begin{cases}
\text{sticking:} \quad \gamma_{\mathrm{t},i}^+ + \epsilon_\mathrm{t}\gamma_{\mathrm{t},i}^- = 0\quad \land & \quad \left|\Lambda_{\mathrm{t},i}\right| < \mu \Lambda_{\mathrm{n},i} \\
\text{slipping:} \quad \gamma_{\mathrm{t},i}^+ + \epsilon_\mathrm{t}\gamma_{\mathrm{t},i}^- \neq 0\quad \land & \quad \Lambda_{\mathrm{t},i} = -\mu \Lambda_{\mathrm{n},i} \operatorname{sign}\left(\gamma_{\mathrm{t},i}^+ + \epsilon_\mathrm{t}\gamma_{\mathrm{t},i}^-\right)
\end{cases}\fp}{newton_tangential}
It can generally be stated that the considered contact laws are commonly used for the modeling of various contact problems in engineering.

\section{Contact kinematics\label{asec:contact_kinematics}}
%
\figc[b]{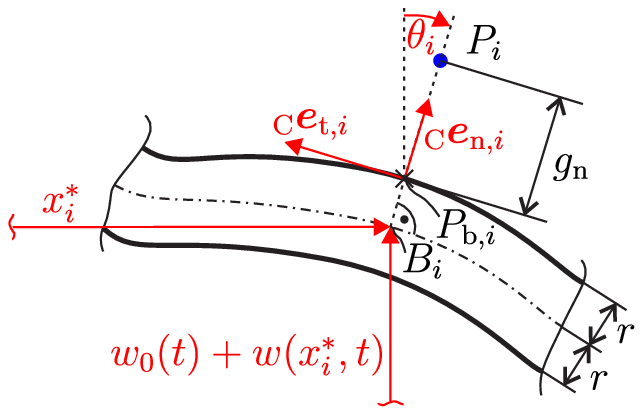}{Contact kinematics}
For each point $P_i$ of the four possible contact points of the slider, a corresponding closest point $P_{\mathrm b,i}$ at the top or bottom surface of the beam can be identified (\cf \fref{contact_kinematics}). This point can be determined by finding the closest location $x_i^*$ on the beam's axis in the Euclidian sense,
\e{x_i^* = \operatorname{argmin}\lbrace\|{}_{\mathrm N}\mms r_{P_i} - {}_{\mathrm N}\mms r_B(x)\|_2^2\rbrace\fp}{closest_point}
Herein, ${}_{\mathrm N}\mms r_B(x) = \matrix{cc}{ x & w_0 + w(x)}\tra$ defines the position vector to the beam's axis (with $w_0 + w(x)$ being the total deflection with respect to the inertial frame of reference), and ${}_{\mathrm N}\mms r_{P_i} = {}_{\mathrm N}\mms r_C + {}^{\mathrm{NS}}\mms A {}_{~\mathrm S}\mms r_{CP_i}$ is the position vector to the point $P_i$. The left subscript notation refers to the associated coordinate system. Three different coordinate systems are defined, namely the inertial coordinate system $\mathrm N$, the contact coordinate system $\mathrm C$ and the slider coordinate system $\mathrm S$ illustrated in \fref{system_and_slider}b. The position vector of the center of mass $C$ in the inertial frame of reference is ${}_{\mathrm N}\mms r_C = \matrix{cc}{x_C & z_C}\tra$. The matrix ${}^{\mathrm{NS}}\mms A$ takes care of the transformation between slider and inertial frame of reference, and it is defined as,
\e{{}^{\mathrm{NS}}\mms A = \matrix{cc}{\cos\phi & \sin\phi\\ -\sin\phi & \cos\phi}\fk}{rotation_matrix_ans}
depending on the slider rotation $\phi$. The contact points $P_1$ through $P_4$ have the coordinate vectors,
\e{{}_{~\mathrm S}\mms r_{CP_1} = \vector{-\frac{b_1}{2}\\ d + \frac{R}{2}},\,
{}_{~\mathrm S}\mms r_{CP_2} = \vector{-\frac{b_1}{2}\\ d - \frac{R}{2}},\,
{}_{~\mathrm S}\mms r_{CP_3} = \vector{+\frac{b_1}{2}\\ d - \frac{R}{2}},\,
{}_{~\mathrm S}\mms r_{CP_4} = \vector{+\frac{b_1}{2}\\ d + \frac{R}{2}}\fk}{rscpi}
in the slider coordinate system.\\
Once the locations $x_i^*$ are found, the contact gap vectors are expressed as
\e{{}_{\mathrm C}\mms g_i = {}^{\mathrm{CN}}\mms A\left({}_{\mathrm{N}}\mms r_{P_i} - {}_{\mathrm{N}}\mms r_{P_{b,i}}\right) = \vector{g_{\mathrm{n},i}\\ g_{\mathrm{t},i}}\fk}{gap}
where, ${}_{\mathrm{N}}\mms r_{P_{b,i}}$ reads
\e{{}_{\mathrm{N}}\mms r_{P_{b,i}} = {}_{\mathrm{N}}\mms r_{B}(x^*_i) + \xi_i\frac{ r}{\sqrt{1+\tan^2\theta}}\vector{-\tan\theta\\1}\fp}{nrpbi}
Herein, $\tan\theta$ is related to the beam's slope at $x^*_i$ by $\tan\theta = w^\prime(x^*_i)$. The parameter $\xi_i$ is $+1$ for $i\in\lbrace 1,4\rbrace$ (top surface), and $-1$ for $i\in\lbrace2,3\rbrace$ (bottom surface). The rotation matrix ${}^{\mathrm{CN}}\mms A$ in \eref{gap} reads,
\e{{}^{\mathrm{CN}}\mms A = \xi_i\matrix{cc}{-\sin\theta & \cos\theta\\ -\cos\theta & -\sin\theta}\fk}{rotation_matrix_acn}
and accounts for the transformation between inertial and contact coordinate systems.\\
The relative velocity ${}_{\mathrm C}\mms \gamma_i = \frac{\dd {}_{\mathrm C}\mms g_i}{\dd t}$ can be derived from \eref{gap}. The result can be split up as
\e{{}_{\mathrm C}\mms \gamma_i = \vector{\gamma_{\mathrm{n},i}\\ \gamma_{\mathrm{t},i}} = \mms W_i\tra\mms u + \mms w_i\fk}{relative_velocity}
where the first term accounts for the contribution of the generalized velocity, and involves the matrix of generalized contact force directions $\mms W_i\tra = \frac{\partial {}_{\mathrm C}\mms g_i}{\partial \mms q\tra}$. The second term in \eref{relative_velocity} accounts for the explicit time dependence of the relative velocity with $\mms w_i = \frac{\partial {}_{\mathrm C}\mms g_i}{\partial t}$. In the considered model, this term stems from the base excitation. The algebraic forms of these derivatives are not presented for brevity.

\section{Time stepping\label{asec:time_stepping}} 
To discretize the measure differential inclusion, we applied the common Moreau time stepping scheme with a constant time step $\Delta t$. Therefore, the generalized coordinates $\mms q^{\mathrm M}$ are first evaluated at the midpoint between the previous and the next time step. This is achieved by taking an Euler step, $\mms q^{\mathrm M} = \mms q^{\mathrm B} + \mms u^{\mathrm B}\frac{\Delta t}{2}$. Quantities at the beginning, the midpoint, and the end of the time step are denoted by a superscript $\mathrm B$, $\mathrm M$ and $\mathrm E$, respectively.\\
The contact kinematics is then evaluated at the midpoint. The set $\mathcal I_{\mathrm C}$ of active contact points is identified as $\mathcal I_{\mathrm C} = \lbrace j\vert g_{\mathrm{n},j}\leq 0\rbrace$. Let $n_{\mathrm C}$ be the number of active contact points. For each active contact point, the kinematical quantities are then evaluated and assembled in the matrix $\mms W = \matrix{ccc}{\mms W_1 & \cdots & \mms W_{n_{\mathrm C}}}$ and the vector $\mms w = \matrix{ccc}{\mms w_1\tra & \cdots & \mms w_{n_{\mathrm C}}\tra}\tra$.\\
In accordance with the Moreau scheme, the velocity vector $\mms u^{\mathrm E}$ at the end of the time step is defined by the second stepping equation $\mms u^{\mathrm E} = \mms u^{\mathrm B} + \mms M^{-1}\left[\mms h\left(\mms q^{\mathrm M},\mms u^{\mathrm M},t^{\mathrm M}\right)\Delta t + \mms W\left(\mms q^{\mathrm M}\right)\mms P\right]$, where $\mms P$ is the unknown contact percussion vector. The contact percussions and the gap velocities need to satisfy the set-valued contact laws. The computation of the compatible contact percussions is addressed in the next subsection. Upon computation of the contact percussions and velocities $\mms u^{\mathrm E}$, the coordinates $\mms q^{\mathrm E}$ at the end of the time step are finally updated using $\mms q^{\mathrm E} = \mms q^{\mathrm M} + \mms u^{\mathrm E}\frac{\Delta t}{2}$.

\subsection{Re-formulation of the contact laws\label{asec:reformulation}}
The set-valued contact laws are enforced by solving an inclusion problem $-\left( \mms \gamma^{\mathrm E} + \mms \epsilon \mms \gamma^{\mathrm B} \right) \in \mathcal N_{\mathcal C}\left(\mms P\right)$, which ensures the compatibility of the contact velocity $\mms \gamma = \matrix{ccc}{\mms\gamma_1\tra & \cdots & \mms\gamma_{n_{\mathrm C}}\tra}\tra$ at the beginning and at the end of the time step, and the contact percussion $\mms P$. This compact formulation contains both contact and impact laws, and involves the matrix $\mms \epsilon = \operatorname{bdiag}\lbrace\mms\epsilon_i\rbrace$ with $\mms\epsilon_i = \operatorname{diag}\lbrace \epsilon_{\mathrm{n},i},\epsilon_{\mathrm{t},i}\rbrace$, and the convex set $\mathcal C = \mathcal S_1 \times \cdots \times \mathcal S_{n_{\mathrm C}}$ with $S_i = \mathbb R_0^+ \times \left[-\mu P_{\mathrm{n},i},\mu P_{\mathrm{n},i}\right]$.\\
By substituting the stepping equation into the inclusion problem, one obtains an implicit algebraic inclusion problem in the unknown contact percussions $\mms P$, %
\e{-\left( \mms G\mms P + \mms c \right) \in \mathcal N_{\mathcal C}\left(\mms P\right)\fk}{inclusion_problem}
with the so-called Delassus matrix $\mms G = \mms W\tra\mms M^{-1}\mms W$ and the vector $\mms c = \left(\mms I + \mms \epsilon\right)\left(\mms W\tra\mms u^{\mathrm B} + \mms w\right) + \mms W\tra\mms M^{-1}\mms h\Delta t$.

\subsection{Computation of the contact percussions\label{asec:augmented_lagrange}}
In the convex analysis framework, the inclusion problem can be re-written as an implicit proximal point equation $\mms P = \operatorname{prox}_{\mathcal C}\left[\mms P - \mms R^{-1}\left(\mms G\mms P + \mms c\right)\right]$ with the metric matrix $\mms R = \mms R\tra > \mms 0$. This corresponds to an augmented Lagrangian approach.
Popular solvers for this equation include the Jacobi and the Gauss-Seidel relaxation method, both of which have been implemented. Both methods involve a sequential application of basic proximal point operations on the subsets of $\mathcal C$. The relevant operations are
\e{\operatorname{prox}_{\mathbb R_0^+}(\xi) = \begin{cases} \xi & \xi\geq 0\\ 0 & \xi<0\end{cases}\fk}{prox_u}
for the unilateral contact and
\e{\operatorname{prox}_{[-a,a]}(\xi) = \begin{cases} a & \xi > a\\ \xi & -a \leq \xi\leq a\\ -a & \xi<-a\end{cases}\fk}{prox_fric}
for the frictional contact.
\end{appendix}


\end{document}